\documentclass[%
 reprint,
superscriptaddress,
 amsmath,amssymb,
 aps,
]{revtex4-1}
\usepackage{graphicx}
\usepackage{amsmath}
\usepackage{float}
\usepackage{comment}
\usepackage{amsmath}
\usepackage{amssymb}
\usepackage{mathrsfs}
\usepackage[below]{placeins}
\usepackage{fancyhdr}
\usepackage{soul,color}

\begin{document}


\title[\texttt{achemso} demonstration]
{Untwinned YBa$_2$Cu$_3$O$_{7-\delta}$ thin films on MgO substrates: a platform to study strain effects on the local orders in cuprates}


\author{Riccardo Arpaia}
\email{riccardo.arpaia@chalmers.se}
\affiliation{Quantum Device Physics Laboratory, Department of Microtechnology and Nanoscience, Chalmers University of Technology, SE-41296 G\"{o}teborg, Sweden}
\affiliation{Dipartimento di Fisica, Politecnico di Milano, Piazza Leonardo da Vinci 32, I-20133 Milano, Italy}
\author{Eric Andersson}
\affiliation{Quantum Device Physics Laboratory, Department of Microtechnology and Nanoscience, Chalmers University of Technology, SE-41296 G\"{o}teborg, Sweden}
\author{Alexei Kalaboukhov}
\affiliation{Quantum Device Physics Laboratory, Department of Microtechnology and Nanoscience, Chalmers University of Technology, SE-41296 G\"{o}teborg, Sweden}
\author{Elsebeth Schr\"{o}der}
\affiliation{Quantum Device Physics Laboratory, Department of Microtechnology and Nanoscience, Chalmers University of Technology, SE-41296 G\"{o}teborg, Sweden}
\author{Edoardo Trabaldo}
\affiliation{Quantum Device Physics Laboratory, Department of Microtechnology and Nanoscience, Chalmers University of Technology, SE-41296 G\"{o}teborg, Sweden}
\author{Regina Ciancio}
\affiliation{CNR-IOM, TASC Laboratory, Area Science Park, Basovizza S.S. 14 km 163.5, I-34149 Trieste, Italy}
\author{Goran Dra\v{z}i\'{c}}
\affiliation{Laboratory for Materials Chemistry, National Institute of Chemistry, Hajdrihova 19, SI-1001 Ljubljana, Slovenia}
\author{Pasquale Orgiani}
\affiliation{CNR-SPIN, University of Salerno, I-84084 Fisciano (SA), Italy}
\author{Thilo Bauch}
\affiliation{Quantum Device Physics Laboratory, Department of Microtechnology and Nanoscience, Chalmers University of Technology, SE-41296 G\"{o}teborg, Sweden}
\author{Floriana Lombardi}
\email{floriana.lombardi@chalmers.se}
\affiliation{Quantum Device Physics Laboratory, Department of Microtechnology and Nanoscience, Chalmers University of Technology, SE-41296 G\"{o}teborg, Sweden}

\date{\today}
\begin{abstract}
We have grown untwinned YBa$_2$Cu$_3$O$_{7-\delta}$ (YBCO) films on (110) MgO substrates that were pre-annealed at high temperature in oxygen atmosphere. The annealing results in surface reconstruction with shallow facets, which induce the suppression of the YBCO twinning domains, and the preferential alignment of the CuO chains along one of the  in-plane directions of the substrate. Because of the large mismatch between the in-plane lattice parameters of film and substrate, the strain induced by the MgO into the YBCO layer is strong and very peculiar. The YBCO film is compressed, with respect to the bulk, and presents a unidirectional buckling of the atomic planes, along the chains' direction, due to a deformation of the copper-oxygen octahedra. The YBCO films, which can be grown with thicknesses down to few unit cells and oxygen doping levels spanning most of the superconducting dome, are patterned into nanowires with dimensions down to 50 nm. The anisotropies due to the untwinning state are preserved in these structures; moreover, additional anisotropies appear, in ultrathin structures where strain effects become more pronounced. Such untwinned and compressively strained films can therefore be used as a platform to study the interplay between strain and the various local orders in the normal state of YBCO.
\end{abstract}

\pacs{}

\maketitle

\section{Introduction}
Strain in thin films can play a crucial role in unveiling the still mysterious phenomenology of the high critical temperature superconductors (HTS) both in the superconducting and in the normal state \cite{keimer2015quantum}. The deformation of the unit cell, with respect to the bulk form, has recently been used as a tool to tune the different orders, characterizing the phase diagram of this class of materials.  Superconductivity \cite{hardy2010enhancement}, electronic nematicity \cite{wu2017spontaneous}, antiferromagnetism \cite{ivashko2019strain} and charge order \cite{bluschke2018stabilization, huang2018modification, kim2018uniaxial}  can all be modified, even strongly, as a consequence of the strain. The response of these orders to  strain might give new insights into the competition and intertwining among the various orders,  possibly shedding light on the mechanisms leading to HTS. Compressive strain can help to settle the long-standing issue about the connection between superconductivity and charge density waves (CDW) in the underdoped region \cite{ghiringhelli2012long, chang2012direct, gerber2015three, comin2016resonant, loret2019intimate}. By applying pressure, the superconducting critical temperature increases \cite{sadewasser2000pressure, cyr2018sensitivity}, while the CDW get strongly modified \cite{huang2018modification, kim2018uniaxial}. These experiments have been performed under conditions, such as hydrostatic pressure and mechanical stress, requiring the use of single crystals.

In principle, it would be advantageous to employ cuprate thin films, where the compressive strain is induced by the substrate. In the case of YBa$_2$Cu$_3$O$_{7-\delta}$ (YBCO), this requires the growth of films which must be untwinned,  to preserve the anisotropies related to the orthorhombicity of the unit cell, and at the same time compressively strained. The growth of untwinned films, where aligned CuO chains are present throughout the sample, is not an easy task since $c$-axis oriented YBCO films usually grow twinned on the substrates commonly used for the deposition of cuprates. This twinning, caused by a random exchange of the in-plane $a$ and $b$ axis, influences the electric and magnetic properties of the films \cite{steinborn1994twinning, schweitzer1996twinning, bernstein2004vortex, frano2016long}. \textcolor{black}{The best untwinned YBCO films, reported so far, have been achieved on SrTiO$_3$ (STO) substrates with a vicinal miscut angle, i.e. with a small misorientation, with respect to the (001) plane direction \cite{PhysRevLett.95.257001, Dekkers2003}}. However, the STO substrate induces a tensile stress into the YBCO films.

Here we report on the growth of untwinned YBCO films under compressive strain. We succeeded in this task, by an in-situ thermal treatment of (110) oriented MgO substrates, prior to the deposition of the YBCO thin films. The thermal treatment results in a reconstruction of the substrate surface, favoring the growth of untwinned YBCO. Moreover, the film is subject to a very peculiar compressive strain, as a result of the large mismatch between the in-plane lattice parameters of the film and the substrate: the $b$ axis shrinks, and tilts around the normal to the MgO surface. 

The manuscript is organized as follows: In  Section \ref{sec: review}, we start with an overview of the past results on  untwinned YBCO thin films, to highlight the crucial conditions which must be fulfilled in order to preserve the alignment of the CuO chains. In Section \ref{sec: MgO} we describe the thermal annealing procedure  performed on the (110) oriented MgO substrates, which induces a surface reconstruction into shallow and elongated facets. Sections \ref{sec: twin} and \ref{sec: strain} describe the structural characterization of our films. Finally, in Section  \ref{sec: wires} we discuss  the transport properties of nano- and micro- wires, patterned on the untwinned YBCO films. The anisotropies related to the orthorhombicity of the YBCO unit cells are preserved down to the nanoscale. In addition, new anisotropies develop, driven by strain. \textcolor{black}{We enphasize here that this growth technique may
not only be relevant for orthorhombic cuprates but also for other superconductors and functional oxides where the in-plane anisotropic properties have to be retained}.

\section{On the quest for untwinned YBCO thin films} \label{sec: review}
Highly textured YBCO thin films are generally deposited by sputtering or by pulsed laser deposition (PLD), at temperatures in the range 700 - 900$^\circ$C. Due to these high temperatures, YBCO films grow in the oxygen depleted tetragonal phase (i.e. with an oxygen content per formula, indicated as $n = 7 - \delta$, in the range $6\le n \le6.4$). During the subsequent cooling, in oxygen atmosphere, the oxygen atoms diffuse into the film, causing a phase transition from tetragonal to orthorhombic. 
 At this step,  a random exchange of the $a$ and $b$ axes can happen, giving rise to the so-called twinned films. In  twinned YBCO films the in-plane anisotropy of the electronic structure and of the local orders, e.g. the charge order, is obscured by twin domains.
All the physical properties, i.e. the cell dimensions and the critical current densities, measured in the two orthogonal in-plane directions, will be averaged on macroscopic scale. Moreover twin planes, separating domains in which $a$ and $b$ directions interchange, act as defects (pinning centers), which weaken locally the superconductivity and can become preferential channels for magnetic-flux penetration \cite{PhysRevLett.67.390}. In untwinned YBCO films, instead, the random exchange of the $a$ and the $b$ axes is eliminated: the crystal structure and the superconducting properties, related to the presence of CuO chains along the $b$ direction, can be singled out  \cite{PhysRevLett.95.257001}. 

\begin{figure}[hbpt!]
\centering
\includegraphics[width=8.5cm]{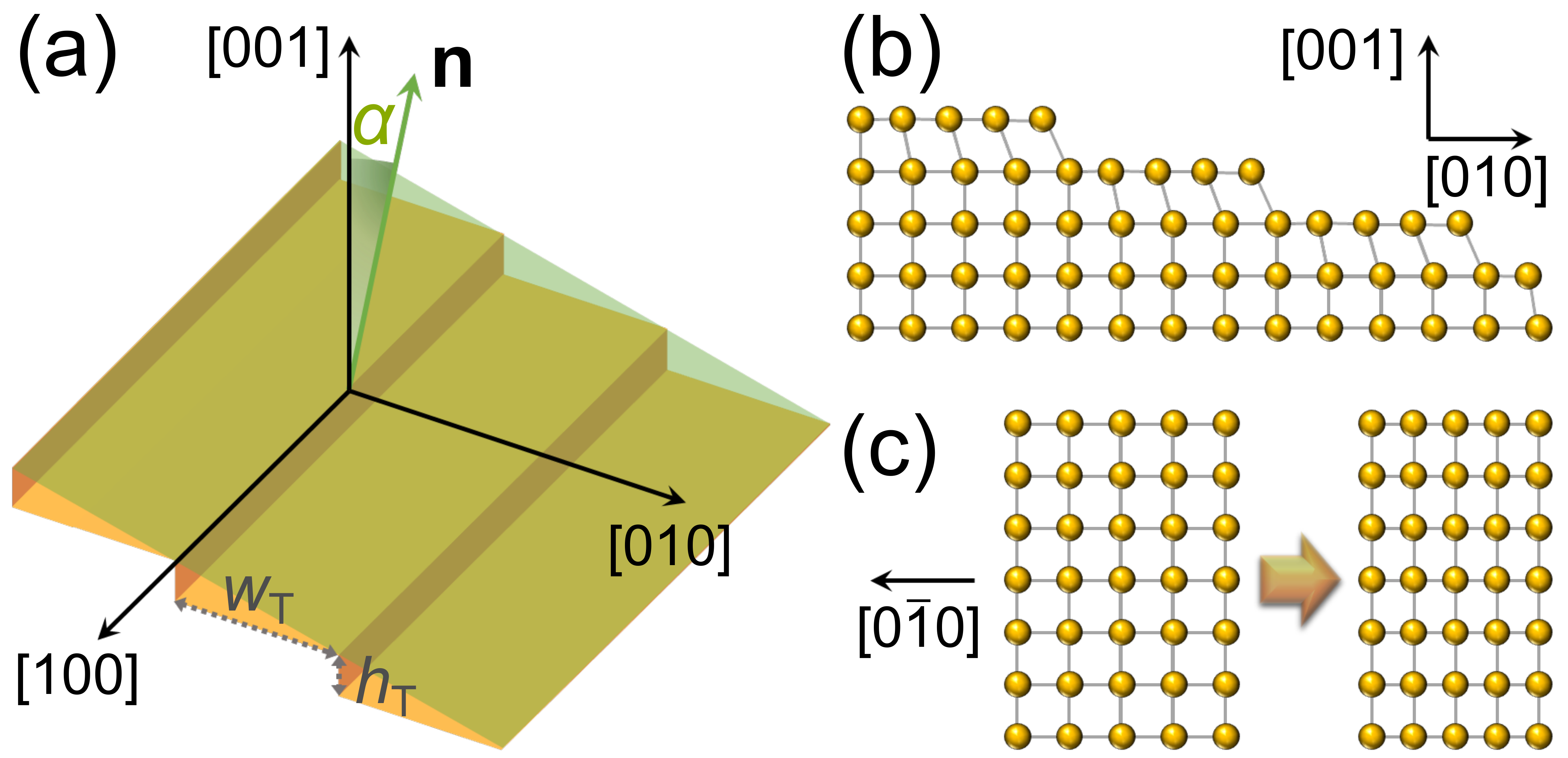}
\caption{(a) A miscut  with  the  surface  normal $\vec{n}$ towards the [010] STO direction leads to (001) terraces with width $w_{\mathrm{T}}$ and height $h_{\mathrm{T}}$. (b) Cross section of the (001) terraces on the STO surface, after a YBCO film is deposited: a compressive strain occurs, perpendicular to the step edges. (c) Top view of a (001) terrace of the STO surface (in the region close to the step), both in its unstrained condition ({\itshape left}) and after a YBCO thin film is deposited on top  ({\itshape right}). Due to the lattice mismatch, the substrate surface is distorted and transforms to a rectangular lattice \textcolor{black}{\cite{zegenhagen1998microscopic, Dekkers2003}}.
} \label{fig:STOSEM}
\end{figure}

The final twinning state of a YBCO film is strongly dependent on the in-plane symmetry of the substrate, because of the mutual strain occurring at the interface.  In particular, it has been shown that substrates characterized by an in-plane lattice that can be transformed into itself \textcolor{black}{either} by a 90$^\circ$ rotation \textcolor{black}{or} by a reflection in a plane perpendicular to the diagonal of the basal plane \textcolor{black}{[i.e. the (110) plane for (001) oriented substrates]} give rise to twinned films \cite{steinborn1994twinning, schweitzer1996twinning}.
\textcolor{black}{This is the case of substrates with a cubic or tetragonal symmetry, such as the (001) oriented MgO, STO and LaAlO$_3$ (LAO), which are the most common substrates used to grow oxides. As a consequence, to modify the twinning state of YBCO, more unconventional substrates have to be chosen.} The best untwinned films achieved so far are on STO substrates with a vicinal angle \cite{PhysRevLett.95.257001, Dekkers2003} \footnote {Partially detwinned films have also been achieved on (001) oriented NdGaO$_3$ substrates. However, the strain induced in YBCO is tensile, as on vicinal angle (001) oriented STO substrates \cite{villard1996ab}. Moreover, 45$^\circ$ grain boundaries and defects,  influencing the transport properties, have been also reported \cite{scherer1992anisotropy, klemenz1999defects}. On (110) oriented NdGaO$_3$ substrates, YBCO films grow instead twinned \cite{steinborn1994twinning, schweitzer1996twinning}. An exception is given by a recent work, where the achievement of untwinned films has been presented \cite{khanof2018coherently}. Here, the strain is tensile along $a$ and compressive along $b$, resulting in a strong suppression of orthorhombicity in the YBCO structure.}. 
Here, the suppression of the twin domains is realized by inducing an \textit{anisotropic strain}, instead of playing with the in-plane symmetry of the substrate \cite{brotz2000detwinning}. Indeed, the surface of STO substrates, having a vicinal angle $\alpha$ towards the  [010] (or the [100]) direction exhibits a periodic sequence of steps, with very straight edges and a height $h_{\mathrm{T}}$, spaced by (001) terraces, whose width $w_{\mathrm{T}}$ decreases with increasing $\alpha$ [see Fig. \ref{fig:STOSEM}(a)]. 
When a thin film is deposited on top of these terraces, a reciprocal strain occurs at the interface between STO and YBCO (which transforms from tetragonal to orthorhombic during the postannealing).  As a consequence of the lattice mismatch between the film and the substrate (see Table \ref{lattice}), the YBCO unit cells are stretched in the plane, while the STO (001) terraces are compressed. 
\begin{table}[!htb]
\begin{tabular}{|c|c|c|}
\hline \hline
surface & in-plane parameters (\AA)\\
\hline
YBCO (001) & $a$ = 3.82; $b$ = 3.89\\
STO (001) & $a$ = $b$ = 3.91\\
MgO (001) & $a$ = $b$ = 4.21\\
MgO (110) & $a$ = 4.21; $b$ = 5.96\\
\hline
\end{tabular}
\caption{\footnotesize{In-plane lattice parameters of the $c$-axis oriented bulk YBCO and of the substrates which are object of investigation in the present work.}}
\label{lattice}
\end{table}
However, the strain on the substrate is strongly anisotropic, since the STO lattice parameter shrinks more easily perpendicular to the step edges, where the atoms are under-coordinated [see Fig. \ref{fig:STOSEM}(b)]. As a consequence, a distortion of the substrate surface occurs at the step: the unstrained square lattice of the (001) terraces transforms into a strained rectangular lattice [see Fig. \ref{fig:STOSEM}(c)]. This distortion has a strong influence on the twinning domains of the YBCO thin films: if $\alpha < 10^{\circ}$ the longer axis parameter of YBCO, $b$, tends to align preferentially along the step edges, where the STO lattice parameter is longer \cite{zegenhagen1998microscopic, brotz2000detwinning}.
By changing the vicinal angle and its orientation, as well as the termination of the STO substrates, different twinning states have been explored in YBCO. In particular, an untwinning degree exceeding 95\% has been reported, using a STO substrate having a vicinal angle $\alpha = 1.1^{\circ}$ towards the  [010] direction \cite{Dekkers2003}.

\vspace{1cm}

\section{Surface reconstruction by temperature annealing of (110) oriented M\MakeLowercase{g}O substrates} \label{sec: MgO}

The reasons for using (110) oriented MgO as substrate to deposit high quality YBCO thin films are manifold. MgO crystals are free from strain defects and twin boundaries, which have been shown (e.g. in LAO) to generate inhomogeneities in the YBCO films \cite{zhuravel2010effect}. They have a thermal expansion coefficient ($\approx 1 \cdot 10^{-5}$ K$^{-1}$ at 273 K) similar to that of YBCO, and are characterized by low dielectric constant ($\approx 10$) and microwave losses (loss tangent $\approx 10^{-5}$ at 10 GHz and 90 K) \cite{mazierska2005precise}. It can therefore be used in devices where a good coupling with the microwave radiation is required, differently from other substrates (e.g. STO), where the high values of dielectric constant and loss tangent, especially at low temperature, are detrimental in such applications. 
Finally, the usage of MgO substrates with the specific orientation (110) has strong implications on the twinning state of YBCO, which have never been explored in previous works, despite the extensive research on this subject \cite{scheel1991substrates, steinborn1994twinning, schweitzer1996twinning, villard1996ab, dubourdieu1996twinning, zegenhagen1998microscopic, brotz2000detwinning, bdikin2002growth, Dekkers2003, nam2005origin, salluzzo2005thickness}. 

The in-plane unit cell of (110) oriented MgO substrates is rectangular ($[001]=4.21$ \r{A}, $[1\bar{1}0]$ $=5.96$ \r{A}). For the symmetry reasons mentioned in the previous section, twin domains should be suppressed, giving rise to untwinned YBCO films. However, this is not the case for (110) MgO, as a consequence of the large mismatch \textcolor{black}{$\delta^{\mathrm{m}}=1-{a_\mathrm{film}}/{a_\mathrm{substrate}}$}  between the in plane lattice parameters of MgO and YBCO ($\delta^{\mathrm{m}}$ is $\approx 9$\% and $\approx 35$\% along the [001] and $[1\bar{1}0]$ MgO directions, respectively), preventing the YBCO unit cell from straining on the substrate. 

For this work we take advantage of the well-known instability of the (110) surfaces of fcc ionic crystals \cite{ho1987stability, gritsch1989mechanism, koch1991new}. Indeed, as a consequence of such instability, these surfaces tend to reorganize into micro and/or nanofacets under certain external conditions (temperature, etching). The consequent mosaic texture can possibly induce a strong in-plane anisotropic strain on the YBCO films, emulating the effect usually obtained with vicinal angle STO substrates. In the case of MgO, the (110) surfaces usually reorganize into (100) oriented facets, which minimize the surface energy, when a chemical polishing of the substrates is performed \cite{henrich1976thermal, de1995atomistic, giese2000atomic}. Alternatively, under thermal annealing the surface energy is reduced through a faceting of the surface into steps of shallow, high Miller index planes \cite{chern1998atomic, sugawara2004faceting}. Even though the dimensions and orientation of the facets  are strongly dependent on the chosen treatment parameters, these elongated domains always run along the [001] MgO direction \cite{chern1998atomic, sugawara2004faceting}.

Therefore, to induce an anisotropic strain to the YBCO thin films, 
we have nanostructured the MgO surface with a heat treatment. In particular, the MgO substrates have been annealed \textcolor{black}{for a time up} to $t_{\mathrm{ann}} = 12$ hours in the deposition chamber, at the same oxygen pressure used for the deposition of the YBCO films ($p_{\mathrm{dep}} = 0.7$ mbar) and at a temperature $T_{\mathrm{ann}} = 790^{\circ}$C, which is slightly higher than the YBCO  deposition temperature. Before the annealing, we have not made any polishing or chemical etching treatment of the surface.

The surface morphology and the orientation preference of the substrate domains have been studied prior to and after the thermal treatment both \textit{in situ} through Reflection High Energy Electron Diffraction (RHEED) patterns and \textit{ex situ} through Atomic Force Microscopy (AFM) images. 

\begin{figure}[hbtp!]
\centering
\includegraphics[width=8.5cm]{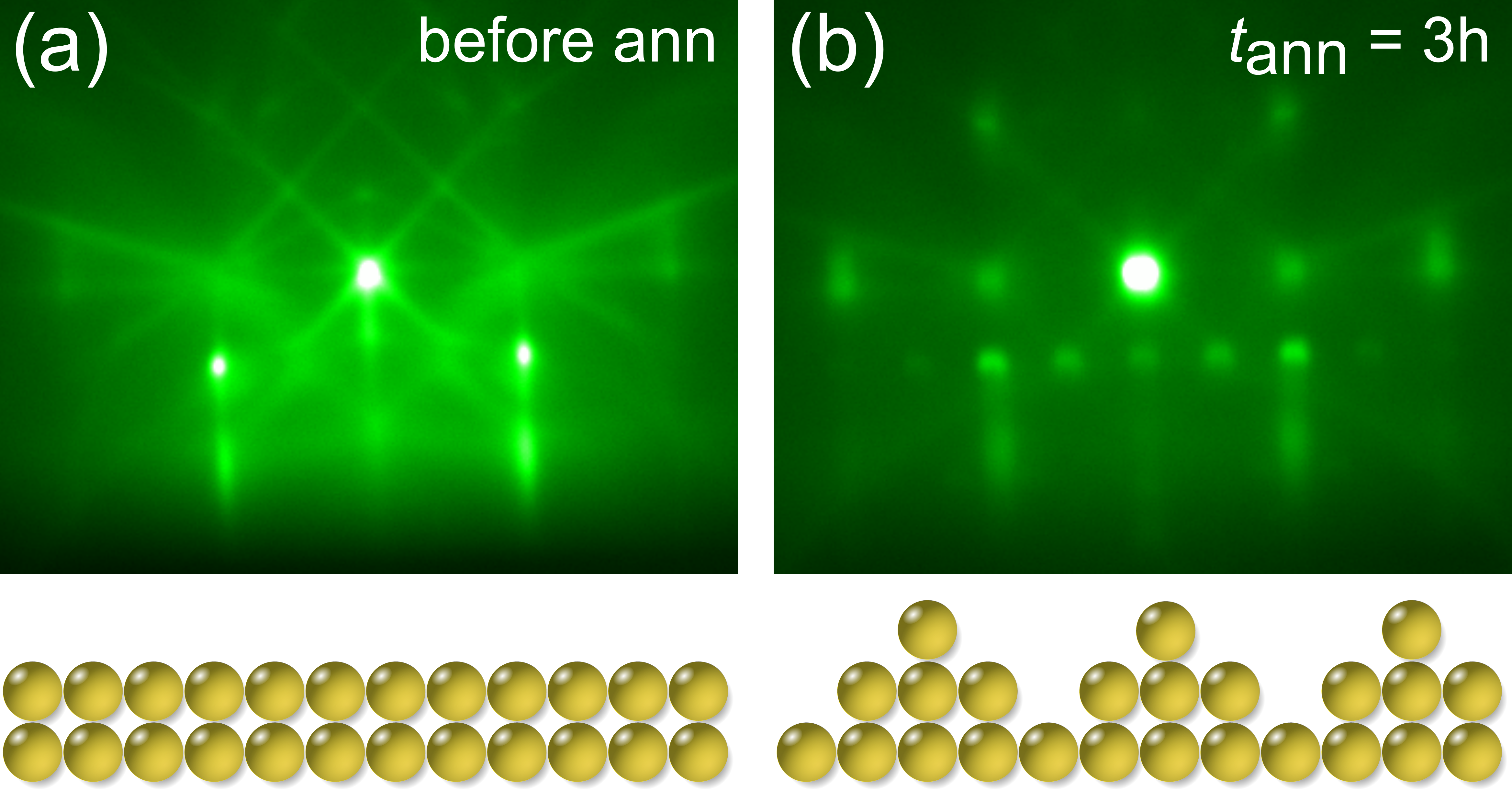}
\caption{{\footnotesize RHEED images of the surface of a (110) oriented MgO substrate (a) before and (b) after a 3 h long annealing done in the PLD deposition chamber at a pressure $p_{\mathrm{dep}} = 0.7$ mbar and at a temperature $T_{\mathrm{ann}} = 790^{\circ}$C. The images have been taken in vacuum and at room temperature.
The incident electron beam is parallel to the $[1\bar{1}0]$ MgO direction.  On  the bottom panel,  cartoons describe   the   surface   morphology   corresponding to the RHEED images: the surface, as a consequence of the annealing, evolves from 2D to 3D.}} \label{fig:RHEED}
\end{figure}

First, we used RHEED to get a qualitative indication of the structure of the MgO surface under thermal annealing (see Fig. \ref{fig:RHEED}). A variation of the surface morphology structure from \textcolor{black}{two-dimensional (2D) to three-dimensional (3D)} is directly revealed by an evolution of the diffraction pattern \cite{ciancio2013evolution}. 
A typical RHEED image of the substrate before any annealing is shown in Fig. \ref{fig:RHEED}(a): a Kikuchi pattern is visible, indicating that the MgO substrate has a pure 2D termination, with clean and smooth surfaces. However, from $t_{\mathrm{ann}} = 2$ hours, the RHEED image evolves from streaky- to spotty-like, and a three-dimensional  diffraction pattern become visible. This pattern, due to the interference of the beam with multiple 3D domains, shows that a robust reconstruction of the MgO surface occurred. After such evolution, the RHEED pattern   remains almost unchanged until   the   end   of  the   annealing procedure [see Fig. \ref{fig:RHEED}(b)].

\begin{figure}[hbtp!]
\centering
\includegraphics[width=8.5cm]{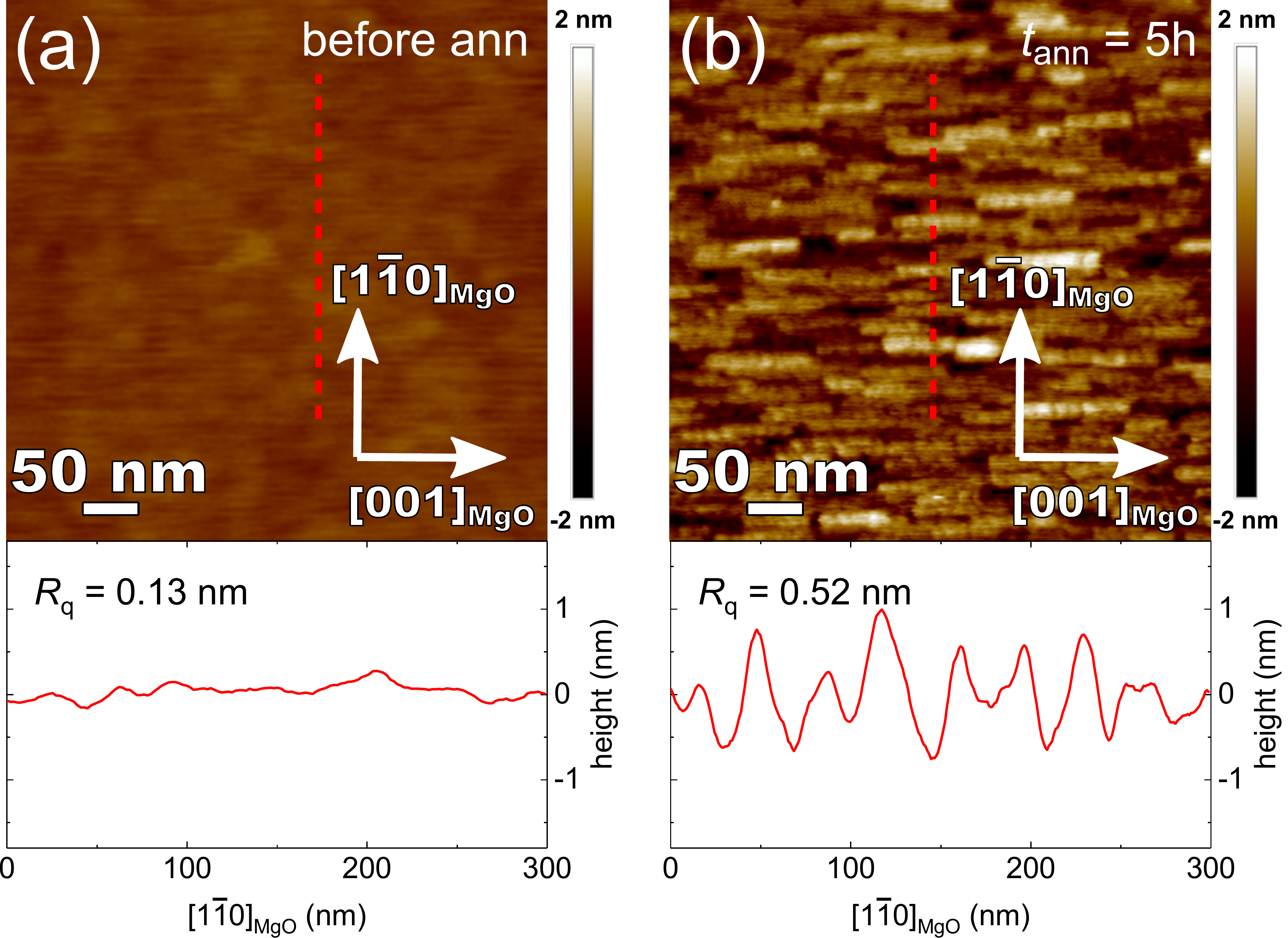}
\caption{{\footnotesize Tapping mode AFM images of the surface of a (110) oriented MgO substrate (a) before and (b) after a 5 h long annealing. As a consequence of the annealing, the surface reconstructs into elongated nanofacets, running along the [001] MgO direction. For each panel, the dashed line represents the direction of the linescan, which is reported in the bottom, together with the value of the roughness $R_{\mathrm{q}}$, given by the standard deviation of the heights of the whole AFM image.}} \label{fig:AFMMgO}
\end{figure}

\begin{figure}[hbtp!]
\centering
\includegraphics[width=8.5cm]{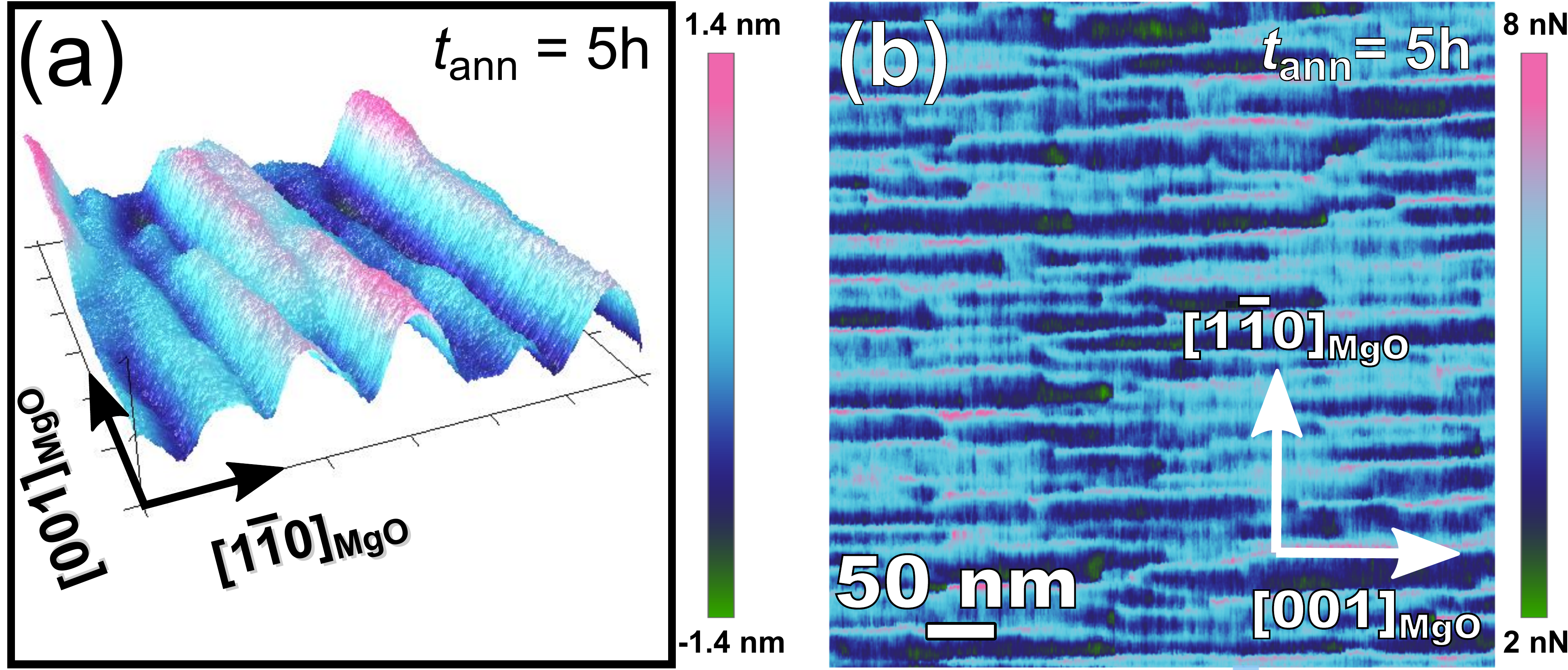}
\caption{{\footnotesize Peak Force Quantitative NanoMechanics (PFQNM) AFM analysis of the surface of a (110) oriented MgO substrate after a 5 h long annealing. (a) PFQNM AFM 3D height image shows elongated nanofacets, compatible with (540)/(450) planes, in agreement with TM AFM results (the unit, defined by the distance between adjacent ticks, is 25 nm in the plane, while it is 1.5 nm out of plane).  (b) PFQNM AFM adhesion image shows the strong chemical contrast due to the different coordination between the atoms  in the valleys between the facets and those along the edges of the facets.}} \label{fig:PFTMgO}
\end{figure}

To obtain more quantitative information about these 3D domains, an AFM investigation has been carried out, using a  Bruker ICON AFM in tapping mode (TM) and Peak Force Quantitative NanoMechanics mode (PFQNM) (see respectively Figs. \ref{fig:AFMMgO} and \ref{fig:PFTMgO}). 
In agreement with RHEED results, the as-delivered substrate surface is reasonably flat, without indication of nanoscale facets [see Fig. \ref{fig:AFMMgO}(a)].
Surface reconstruction occurs instead after the annealing [see Fig. \ref{fig:AFMMgO}(b)]: the surface roughness of the film increases by a factor 4, and elongated islands, running along the [001]  in-plane direction, are formed. The average length of these islands is $\approx 80$ nm, while width and height vary in the range $25-30$ nm and $1-1.6$ nm respectively. As confirmed by the TM AFM linescans, the angle between the plane of the facets and the (110) of the substrate is in the range 3.5$^{\circ}$-7$^{\circ}$. Similar facet dimensions and slopes have been previously observed in literature \cite{sugawara2004faceting} and attributed to limitations of the TM AFM, due to finite tip radius and large tip-surface distance. In order to overcome these limitations, we have additionally performed an investigation with PFQNM AFM (see Fig. \ref{fig:PFTMgO}). 
In Peak Force mode, the peak of force curves at each pixel is used as the imaging feedback signal. The tip-surface distance in the PFQNM AFM is similar to contact AFM, thus enhancing the lateral resolution of imaging. In addition, PFQNM AFM allows quantitative determination of nano-mechanical properties of the surface from analysis of force curves. The tip used in the PFQNM measurements has nominal radius of 2 nm. The measured topography [see Fig. \ref{fig:PFTMgO}(a)] confirmed the facet dimensions and slope that have already emerged in TM AFM. Adhesion images [see Fig. \ref{fig:PFTMgO}(b)] show a strong contrast along the $[1\bar{1}0]$ MgO direction (i.e. perpendicular to the islands elongation), between atoms on the top of the facets and those on the valleys between facets. The reason for this contrast is related to a gradient of coordination along the $[1\bar{1}0]$ direction: the atoms in the valleys between the facets are over-coordinated, while those along the edges of the facets are under-coordinated. Consequently, along the $[1\bar{1}0]$ MgO direction the atoms are more prone to be strained at the interface with the film, similarly to what occurs along the $[0\bar{1}0]$ direction of vicinal angle STO substrates [see Fig. \ref{fig:STOSEM}(b)-(c)].

The sample topography, as determined via AFM investigation, implies that the surface of a (110) oriented MgO substrate reconstructs - as a consequence of our annealing procedure - into shallow facets, formed by (540)/(450) planes, or similar high Miller index planes. To justify the presence of nanometer-sized facets with such orientation, which has never been taken into account in previous reports, we have calculated the surface energy of (540)/(450) facets. The calculations were carried out using density functional theory (DFT) with the van der Waals density functional (vdW-DF) method \cite{dion2004van, thonhauser2007van, lee2010higher, berland2015van} in the consistent exchange version \cite{berland2014exchange}. In particular, we used the DFT code Quantum Espresso \cite{giannozzi2009quantum} with a fast-Fourier-transform implementation of the central integral in the nonlocal correlation calculations \cite{roman2009efficient}. The objects of the calculations are slabs of MgO with different surface termination, having thickness of $\approx 1.5$ nm. In particular, the surface energy of the reference slab of (110) oriented MgO has been compared to that of its faceted versions. The calculated surface energy $E_{\mathrm{s}}$, defined as the energy cost for creating the surface by cleaving a piece of bulk material and letting the atomic positions relax, is reported per surface area in Table \ref{Elsebeth}.
\begin{table}[!htb]
\begin{tabular}{c|c|c}
\hline \hline
surface & angle & $E_{\mathrm{s}}$ (eV/\AA$^2$)\\
\hline
MgO (110) & 0$^{\circ}$ & 0.154 \\
MgO (450) & 6.3$^{\circ}$ &  0.142 \\
MgO (350) & 14$^{\circ}$ & 0.126 \\
MgO (100) & 45$^{\circ}$ & 0.072 \\
\hline
\end{tabular}
\caption{\footnotesize{For each surface termination, the surface energy $E_{\mathrm{s}}$ per surface area, determined via DFT calculations, is reported together with the angle between the facets and the as-delivered MgO substrate orientation. In particular, the $E_{\mathrm{s}}$ of the relaxed (110) MgO has been compared with that of the  (450) facets measured by AFM. For reference, we have also calculated $E_{\mathrm{s}}$ for the (350) and (100) facet terminations, which have been reported in previous works on (110) MgO substrates under different thermal and chemical treatments \cite{henrich1976thermal, de1995atomistic, giese2000atomic, chern1998atomic}.}}
\label{Elsebeth}
\end{table}
The result of the calculations is that the surface energy of the relaxed MgO (110) surface is reduced by the creation of the facets, despite the small energy cost required in the latter case for the creation of the cusps. As expected for MgO, the surface energy is minimized by introducing (100) facets. However, already the reconstruction into high Miller index (540)/(450) facets, that we have observed via AFM as a consequence of the thermal treatment, is energetically more favorable than the (110) surface. The (540)/(450) facet reconstruction implies less atom rearrangement compared to the (100) and (350) facets, which can explain why it prevails.

\section{Dependence of the YBCO twinning state on the MgO preannealing time} \label{sec: twin}

We have deposited YBCO films on (110) oriented MgO substrates by pulsed laser deposition ($T_{\mathrm{dep}} = 760^{\circ}$C, $p_{\mathrm{dep}} = 0.7$ mbar, laser fluence 1.5 J/cm$^2$ \cite{arpaia2016yba2cu3o7}) under two different conditions: soon after the YBCO deposition temperature is reached (i.e. without preannealing of the substrate), and after the MgO preannealing, for a time which ranged between $t_{\mathrm{ann}} = 1-12$ hours. In this section, we will focus on slightly overdoped \footnote {Slightly overdoped films are the best systems where to study the twinning state, since at this oxygen doping level the in-plane orthorhombicity of the YBCO unit cell, i.e. the difference in length between the $a$ and the $b$ axes, is strongest \cite{ye1993quantitative}.}, 50 nm thick YBCO films, achieved by using a postannealing oxygen pressure of 900 mbar \cite{baghdadi2015toward}. However, the same discussion can be extended to films with thickness down to few unit cells and with oxygen doping spanning down to the strongly underdoped region of the phase diagram, whose general properties have been described in previous works \cite{arpaia2017transport, arpaia2018probing}.

AFM shows that, independently of the preannealing time of the substrate, all grown YBCO films have smooth surfaces with an average roughness of one atomic cell. There are some holes that coexist with the 3D spiral-like  domains, typical of a $c$-axis growth. Additional details on the morphological properties of the films can be found in Refs.  \cite{arpaia2017transport, arpaia2018probing}. 

\begin{figure}[hbtp!]
\centering
\includegraphics[width=8.5cm]{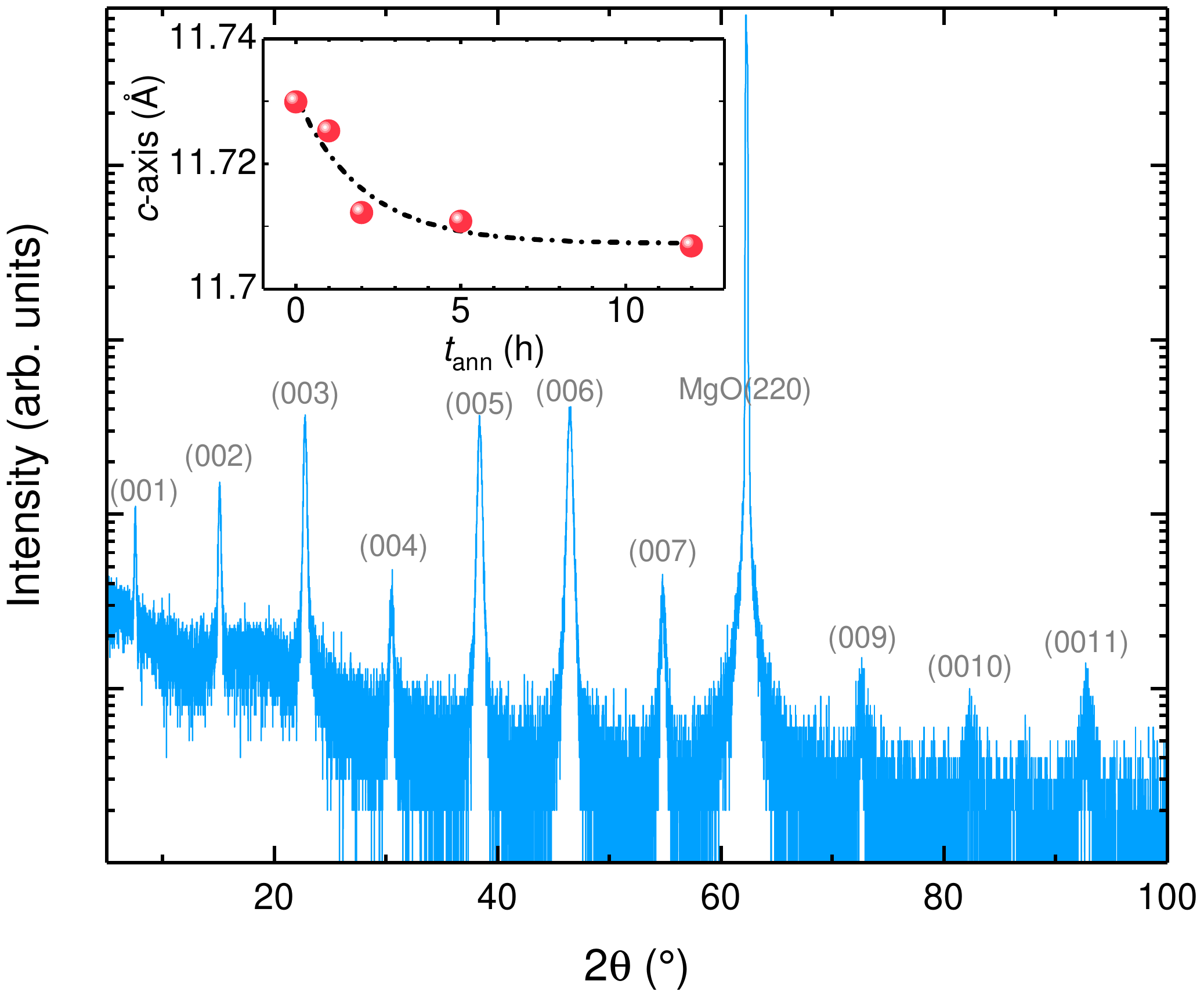}
\caption{{\footnotesize  XRD $2\theta-\omega$ scan of a YBCO film (thickness $t = 50$ nm) on a (110) oriented preannealed ($t_{\mathrm{ann}} = 2$ hours) MgO substrate. Only (00$\ell$) YBCO reflections are observed, from whose position a compressive in-plane strain induced by the substrate into the films can be inferred. In the inset, the $c$-axis peak extracted by the (00$n$) reflections is shown, and plotted as a function of the preannealing time of the substrate. The dashed line is a guide for the eye.}} \label{fig:MgOXRDth}
\end{figure}

The structural properties have been determined by X-Ray Diffraction (XRD) analysis using a Panalytical X'Pert PRO Materials Research 4-axis diffractometer with 4-bounce Ge220 monochromator and PIXcel detector. Symmetric $2\theta-\omega$ scans confirm that the films are highly crystalline and $c$-axis oriented (see Fig. \ref{fig:MgOXRDth}). 
\begin{figure*}[t]
\centering
\includegraphics[width=17.7cm]{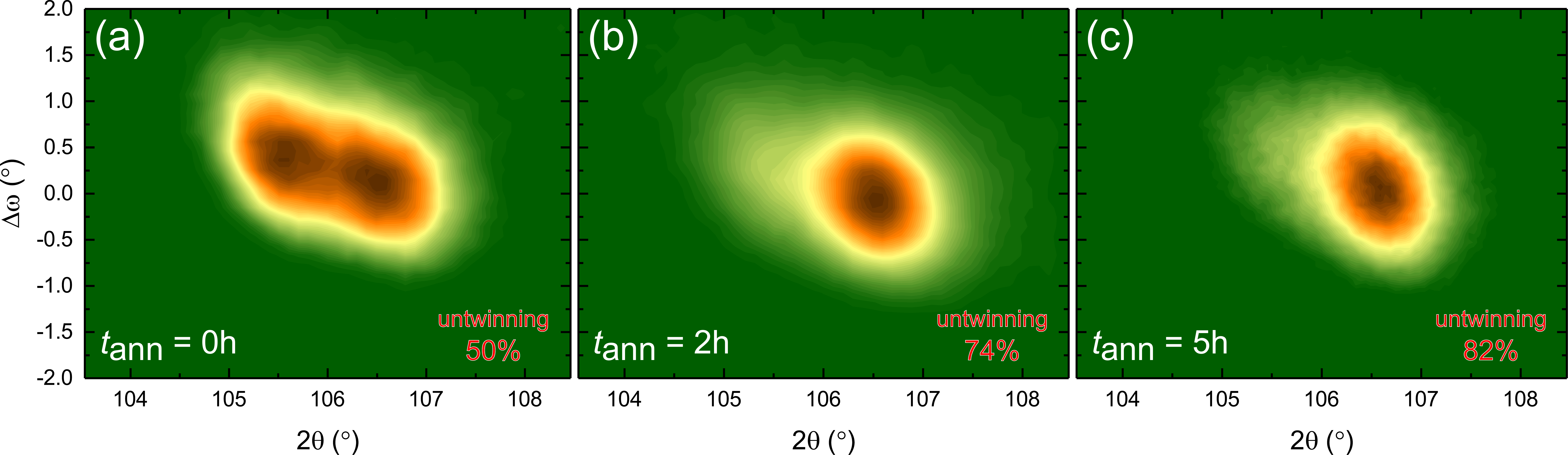}
\caption{{\footnotesize XRD $2\theta-\omega$ maps, along the $[1\bar{1}0]$ MgO direction, of the (038) and (308) reflections for YBCO films deposited on MgO (a) with $t_{\mathrm{ann}}=0$ h; (b) with $t_{\mathrm{ann}}=2$ h; (c) with $t_{\mathrm{ann}}=5$ h. The orientation relationships we infer from the maps of the untwinned films, as in panels (b) and (c), are YBCO$[100]||$MgO$[1\bar{1}0]$ and YBCO$[010]||$MgO$[001]$.}} \label{fig:MgOXRDmap}
\end{figure*}
From the angular position of the (00$\ell$) peaks we can estimate the length of the YBCO $c$-axis parameter. It is plotted as a function of the MgO preannealing time in the inset of Fig. \ref{fig:MgOXRDth}. Without preannealing, the $c$-axis is 11.73 \r{A} long. When the MgO preannealing time is increased, the extracted length of the $c$-axis parameter decreases: after $t_{\mathrm{ann}} = 12$ hours, the $c$-axis length is 11.71 \r{A}. However, the extracted value of the $c$-axis saturates for preannealing times longer than 2 hours, in agreement with the complete reconstruction of the MgO surface observed by RHEED. The result shown in Fig. \ref{fig:MgOXRDth} is a first demonstration that the reconstruction of the MgO substrate, induced by our thermal treatment, has a clear influence on the structure of the YBCO films. The important role the MgO \textcolor{black}{itself} plays in defining the properties of YBCO is also highlighted by the value of  the $c$-axis parameter, which - at any preannealing time - is elongated with respect to the bulk value ($c_\mathrm{bulk} = 11.69$ \AA). This occurrence indicates a  \textit{compressive in-plane strain} induced by the substrate.


To determine the twinning state of the YBCO films as a function of the MgO preannealing time, and the film-substrate orientation relations, we have explored by XRD $2\theta-\omega$ maps the asymmetrical (038)-(308) reflections of YBCO. In relaxed films these two reflections are at $2\theta=105.31^\circ$ and  $2\theta=106.80^\circ$ respectively, therefore the splitting of the reflections associated to $a$ and $b$ axes can be resolved. For each film the maps have been taken both along  the $[001]$ and the $[1\bar{1}0]$ MgO directions. In Fig.  \ref{fig:MgOXRDmap}, the results along the $[1\bar{1}0]$ direction are presented for films deposited on differently annealed substrates.

For films deposited on substrates without annealing [see Fig. \ref{fig:MgOXRDmap}(a)] both the (038) and the (308) reflections are present in the map with the same intensity. The untwinning degree is determined by the integral of the stronger peak compared to the total integral of the two peaks. It  corresponds to the percentage of the most frequent YBCO in-plane parameter along the probed MgO in-plane orientation. For substrates without annealing we get an untwinning degree of 50\%.   The film is therefore perfectly twinned, since a random exchange of the $a$-axis (represented by the presence of the (308) reflection) and $b$-axis (represented by the presence of the (038) reflection) is present. 

A strikingly different result is obtained in films deposited on annealed substrates. Here, along the $[1\bar{1}0]$ MgO direction, the (308) peak ($a$ axis) is predominant, while along the  $[001]$ MgO direction the (038) reflection ($b$ axis) is the strongest. Such result, which represents one of the main achievements of this work, is shown in Figs. \ref{fig:MgOXRDmap}(b) and \ref{fig:MgOXRDmap}(c) for YBCO films deposited on MgO substrates annealed respectively 2 and 5 hours. The untwinning degree of YBCO is 74\% after $t_{\mathrm{ann}}=2$ h, while it saturates to 82\% after $t_{\mathrm{ann}}=5$ h, a value close to those obtained in YBCO films deposited on vicinal angle STO substrates \cite{Dekkers2003}. 

The presence of [001] oriented elongated facets on the surface of annealed MgO substrates favors the growth of almost untwinned YBCO films. The $a$ axis tends to align along the $[1\bar{1}0]$ MgO direction, i.e. perpendicularly to the facets, where the substrate atoms are under-coordinated and more prone to be strained at the interface with the film; the CuO chains ($b$ axis) are instead oriented along the facets. The film-substrate orientation relationship has severe implications on the way the strain is applied to the film. This is indeed confirmed by the length of the in-plane YBCO lattice parameters we infer from the position of the (038) and (308) reflections [see Fig. \ref{fig:MgOXRDmap}(c)]. The $a$ axis is 3.82 \AA, very close to the relaxed bulk value. On the contrary, all the strain induced by the substrate is on the $b$ axis, which is compressed down to a value of 3.87 \AA (from 3.89 \AA).

\section{A peculiar strain: unidirectional buckling of the YBCO atomic planes} \label{sec: strain}

The XRD investigation reveals additional details of the peculiar strain relaxation behavior of our films. 
\begin{figure*}[t]
\centering
\includegraphics[width=17.7cm]{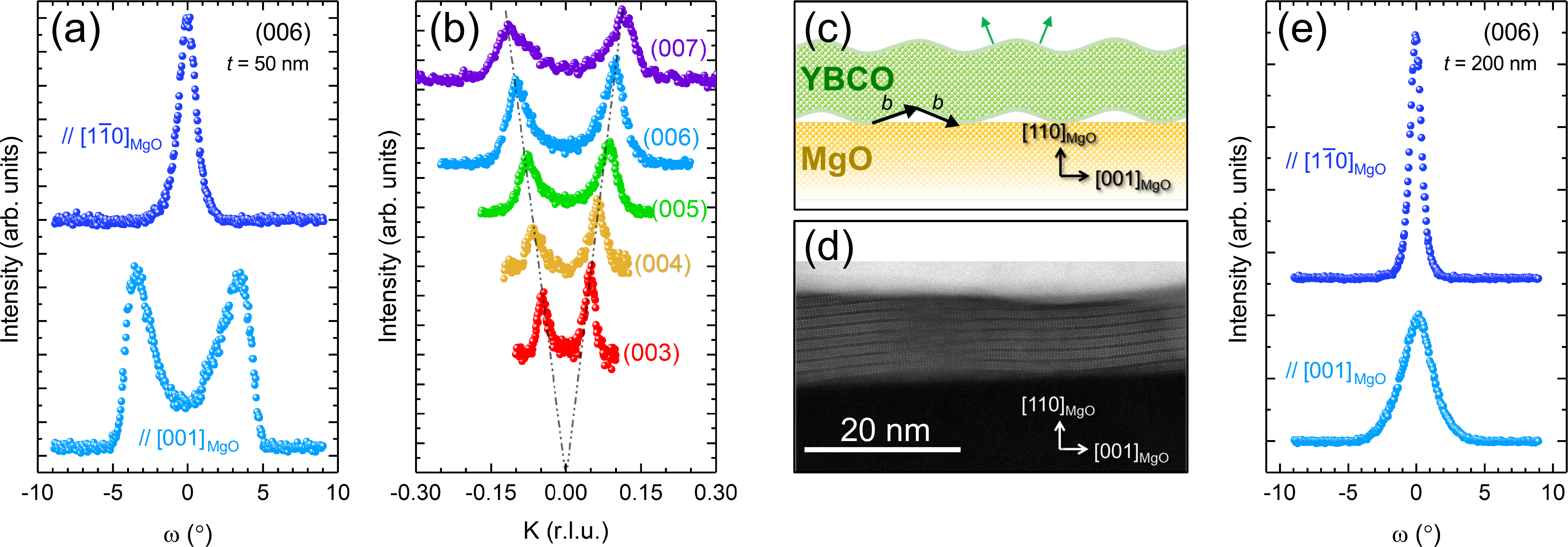}
\caption{{\footnotesize (a) XRD $\omega$ scan of the (006) YBCO reflection, measured along the $[001]$  and $[1\bar{1}0]$ MgO directions, for a 50 nm thick film. The splitting along the MgO $[001]$ highlights a tilt of the (001) YBCO planes occurring along the [010] YBCO direction. (b) XRD transverse $K$ scans of several (00$\ell$) YBCO Bragg peaks (the scans are in units of the YBCO reciprocal lattice vector $b^* = 2\pi/b$, \textcolor{black}{and they have been shifted linearly along the vertical direction}). The splitting increases linearly with the order of the reflection, pointing toward the existence of macroscopic buckling domains. (c) Sketch of the buckling of the $ab$ planes occurring along the $b$ axis direction of YBCO films grown on (110) MgO substrates. (d) Cross sectional HAADF-STEM image  of a 10 nm thick YBCO film, taken in the [010] zone axis, enlightens the presence of a waving of the (001) YBCO planes.  (e) Same as panel (a), but here the film thickness is 200 nm. Above a certain threshold in thickness, the film relaxes and the unidirectional buckling of the atomic planes disappears.}} \label{fig:Buckling}
\end{figure*}
Figure \ref{fig:Buckling}(a) shows the rocking curve of a typical film, measured along \textcolor{black}{both main in-plane directions}. Along the $[1\bar{1}0]$ MgO  direction we get a single peak, aligned to the normal direction of the substrate ($\omega =$ 0), with a full width at half maximum (FWHM) of $\approx 1.5^{\circ}$ which supports highly textured thin films.
Instead, along the $[001]$ MgO direction a much broader (00$\ell$) YBCO reflection is present, with two maxima having similar intensities, and an $\omega$ offset of $\approx 3^{\circ}$ with respect to the substrate normal. Such occurrence is associated to a buckling of the $ab$ atomic planes, which is a well known phenomenon in perovskite films when they are grown on substrates with large lattice mismatch, as in our case \cite{woodward1997octahedral1, woodward1997octahedral2}. Indeed in manganite \cite{gebhardt2007formation, vailionis2009anisotropic}, ruthenate \cite{vailionis2011misfit} and nickelate \cite{weber2016multiple} films the distortion of the oxygen octahedra represents the main strain relaxation mechanism, requiring a lower energy cost than the compression/stretching of the in-plane bond lengths. In particular, by plotting  in the reciprocal space the rocking curves of the different (00$\ell$) YBCO reflections, measured along the $[001]$ MgO direction, we observe that the distance between the two peaks increases with the order of the reflection [see Fig. \ref{fig:Buckling}(b)]. This indicates that our films are dominated by macroscopic buckling domains \cite{gebhardt2007formation}, with the (001) atomic planes tilted by an angle $\approx 3^{\circ}$, with respect to the substrate surface, as sketched in Fig. \ref{fig:Buckling}(c). These buckling domains are unidirectional, i.e. oriented along the $[001]$ MgO direction, which - if the substrates are preannealed - corresponds to the direction of the CuO chains of YBCO. \textcolor{black}{They are a pure strain effect, without a direct connection with the presence of the ``waves'' observed in the reconstructed surface of the substrates, which run along the perpendicular direction}.


To understand the origin of the buckling in our films, we performed high-angle annular dark-field scanning
transmission electron microscopy (HAADF-STEM). A representative Z-contrast image is shown in Fig. \ref{fig:Buckling}(d). A waving of the (001) YBCO planes along the $[001]$ MgO direction is enlightened, which is also favored by the intercalation of Y$_2$Ba$_4$Cu$_8$O$_{16}$ intergrowths (e.g. YBCO unit cells with a doubling of the
chains layers) within the YBCO matrix \cite{arpaia2017transport}. The average period of the waves is $\approx 30$ nm.
Since this length is commensurate to several tens ($\approx 70$) of MgO
unit cells, strain relaxation of the film occurs with a slight tilt of the (001) YBCO planes around
the normal to the substrate surface. This ensures the
absence of grain boundaries at the interface between adjacent
buckling domains \cite{chisholm1991structural, lombardi1998transport}, which is also confirmed
by the transport characterization on our YBCO films,
presented in the next section.

The \textcolor{black}{aforementioned} HAADF-STEM investigation has been performed on ultrathin YBCO films ($t = 10$ nm), where interface effects between film and substrate are emphasized. Indeed, the buckling of the (001) YBCO atomic planes is a mechanism to release the stress at the interface with the substrate, when a large lattice mismatch is present. We expect this to occur up to a relaxation threshold, i.e. a thickness above which the film relaxes, growing independently of the substrate lattice parameters. On 200 nm thick YBCO films the transverse splitting of the (00$\ell$) Bragg peaks along the $[001]$ MgO direction disappears  [see Fig. \ref{fig:Buckling}(e)]. The peak is still broader than on the  $[1\bar{1}0]$ MgO direction, as a consequence of the buckling occurring in the first layers, closer to the interface.

\section{Anisotropies in the transport properties of untwinned YBCO nanowires} \label{sec: wires}
For our YBCO films grown on (110) oriented MgO substrates we have developed a nanopatterning procedure based on a PLD-deposited carbon mask, in combination with electron beam lithography and a gentle Ar$^+$ ion milling \cite{arpaia2013improved, baghdadi2015fabricating}. We have therefore fabricated nanowires \cite{nawaz2013microwave, arpaia2017transport, arpaia2018probing}, Dayem bridge Superconducting QUantum Interference Devices \cite{arpaia2014ultra, arzeo2016toward, arpaia2016improved, trabaldo2019transport} and nanorings \cite{arpaia2014yba2cu3o7} with dimensions down to 30 nm, \textcolor{black}{finding that the properties are unaffected by the patterning procedure}.

\textcolor{black}{Both untwinned and twinned samples can be patterned using the same nanofabrication routines.} We can therefore focus on the transport properties of untwinned nanostructures, representing clean systems where the alignment of the CuO chains is preserved down to the nanoscale. On these systems, we can measure both an intrinsic anisotropy, deriving \textcolor{black}{from} the orthorhombicity of the unit cell (i.e. by the presence of CuO chains along the $b$ direction), and possibly unexpected anisotropies, appearing as a function of the oxygen doping and/or of the film thickness as a consequence of strain and confinement effects. 

\begin{figure}[hbtp!]
\centering
\includegraphics[width=8.5cm]{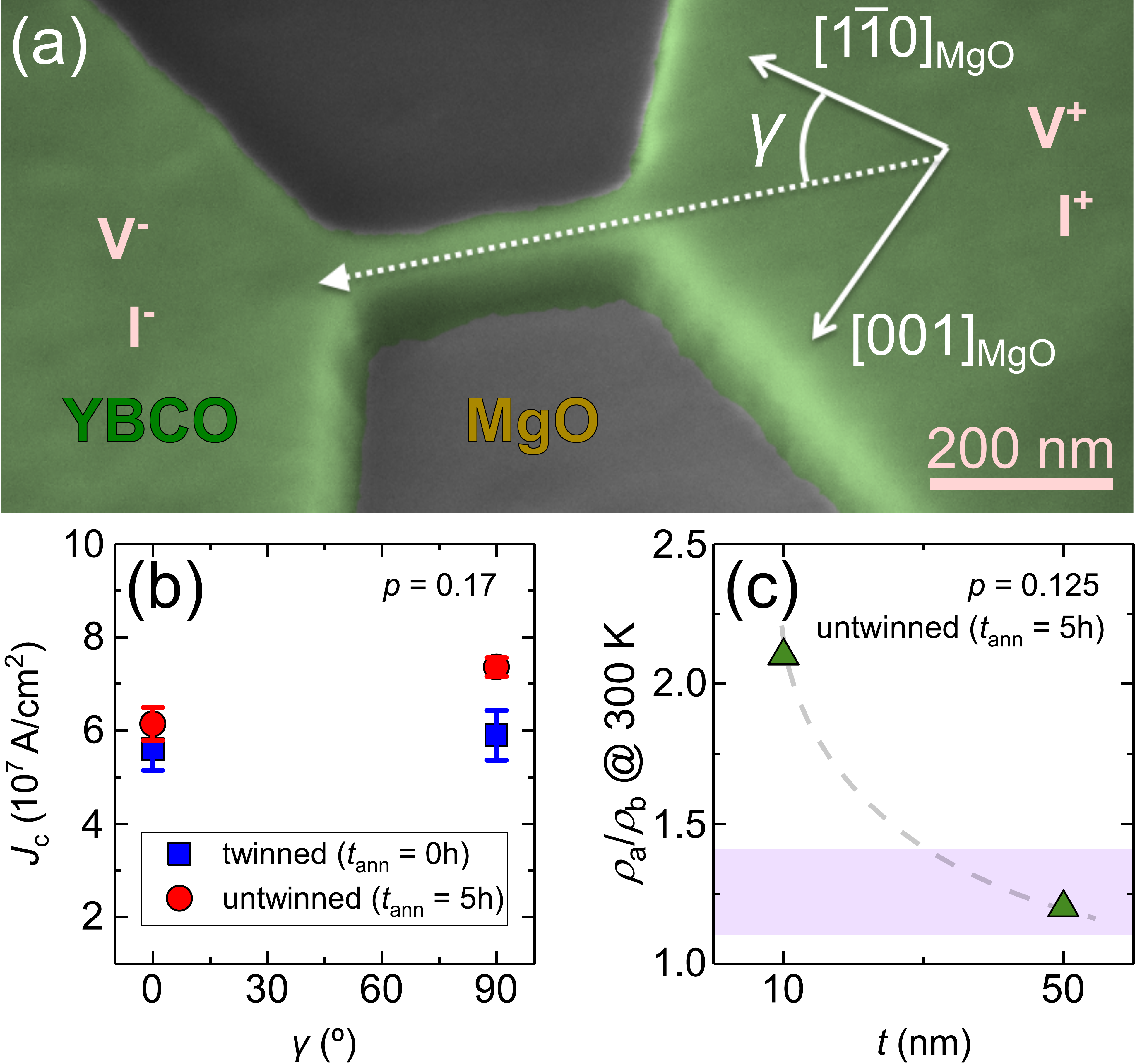}
\caption{{\footnotesize (a) Tilted Scanning Electron Microscope picture of a 50 nm wide YBCO nanowire, patterned at an angle $\gamma$ with respect to the $[1\bar{1}0]$ MgO  direction. (b) Critical current density $J_{\mathrm{c}}$ measured at 4.2 K both in twinned (squares) and in untwinned (circles) slightly overdoped, \textcolor{black}{50 nm thick} nanowires, patterned at $\gamma = 0^\circ$ and $\gamma = 90^\circ$. For each angle we have measured a large number of nanowires,  determining a distribution of $J_{\mathrm{c}}$ values. Each data point and error bar in the plot represents the average and the variance of the distribution, respectively. Only for untwinned nanowires we have a $J_{\mathrm{c}}$ anisotropy, since at $\gamma = 90^\circ$ ($b$ axis) the \textcolor{black}{critical} current is a factor 1.2 larger than at $\gamma = 0^\circ$ ($a$ axis). (c) The ratio of the resistivities $\rho_{\mathrm{a}}/\rho_{\mathrm{b}}$, measured at 300 K on untwinned underdoped nanowires, is presented as a function of the film thickness $t$. The solid area represents the range of $\rho$ anisotropy, which is expected in single crystals at that level of doping \cite{ando2002electrical}. The dashed line is a guide for the eye. In ultrathin 10 nm thick nanowires an additional, strain-induced, anisotropy is present.}} \label{fig:Nanowires}
\end{figure}

To study the in-plane anisotropy of the films due to their twinning state, and test if it is preserved down to the nanoscale, we have patterned YBCO nanowires on twinned and untwinned films [with untwinning degrees respectively as in Figs. \ref{fig:MgOXRDmap}(a) and \ref{fig:MgOXRDmap}(c)] at different in-plane angles 
$\gamma$ with respect to the $[1\bar{1}0]$ direction of the substrate [see Fig. \ref{fig:Nanowires}(a)].
Among all the measurements which have been done, in wires of different widths (between 50 nm and 5 $\mu$m) and oxygen levels (between the strongly underdoped and the slightly overdoped  \cite{arpaia2018probing}), here we will focus on the values at $\gamma = 0^\circ$ and $\gamma = 90^\circ$ of the critical current density $J_{\mathrm{c}}$ at 4.2 K and of the resistivity $\rho$ at 300 K. These values have been respectively determined by recording the Current Voltage Characteristics and measuring the resistance as a function of the temperature.

In Fig. \ref{fig:Nanowires}(b) the $J_{\mathrm{c}}$ values at 4.2 K of slightly overdoped (hole doping $p = 0.17$ \textcolor{black}{\footnote{The hole doping $p$, number of holes for planar copper atom, has been determined in our films indirectly, combining the knowledge of the $c$-axis parameter, estimated via XRD with that of the critical temperature $T_c$, determined via resistance vs temperature measurements. A detailed description of the analysis we have applied is presented in Ref. \cite{arpaia2018probing}.}}) nanowires, \textcolor{black}{with thickness of 50 nm, and width in the range between 50 and 300 nm}, are shown at $\gamma = 0^\circ$ and $\gamma = 90^\circ$. \textcolor{black}{On the untwinned film, nanowires patterned at $\gamma = 90^\circ$ (along the YBCO $b$-axis) have the  $J_{\mathrm{c}}$ values which are a factor 1.2 larger than at $\gamma = 0^\circ$ (nanowires patterned along the YBCO $a$-axis).} The reason of such anisotropy is intimately related to the twinning state of the systems. Indeed, the $J_{\mathrm{c}}$ values of our nanowires \textcolor{black}{approach the depairing critical current value, in the limit of Abrikosov vortices entry \cite{arpaia2014resistive, charpentier2016hot, arpaia2018probing}.} Here the anisotropy of the critical current density $J^b_{\mathrm{c}}/J^a_{\mathrm{c}}$ is given by the anisotropy of the London penetration depth $\lambda^a_{\mathrm{L}}/\lambda^b_{\mathrm{L}}$ in the $ab$ planes of YBCO, related to the presence of CuO chains along the $b$ direction \cite{baghdadi2017study}. The $\lambda_{\mathrm{L}}$ in-plane anisotropy inferred from the $J_{\mathrm{c}}$ values of our nanowires is in \textcolor{black}{fairly} good agreement with that determined in untwinned YBCO single crystals with the same value of doping \cite{basov1995plane, ishida1997two}. \textcolor{black}{The small discrepancy of the value we have estimated with respect to the highest anisotropy value reported in Ref. \cite{basov1995plane} ($\lambda^a_{\mathrm{L}}/\lambda^b_{\mathrm{L}} = 1.6$) can be explained consirering the non-complete detwinning of our thin films.} On the contrary, for nanowires patterned on twinned films the $J_{\mathrm{c}}$ values are almost identical at $\gamma = 0^\circ$ and $\gamma = 90^\circ$. Indeed here, because of the random exchange of the $a$ and $b$ axis, the effect of the $\lambda_{\mathrm{L}}$ anisotropy due to the cell orthorhombicity is lost.

In Fig. \ref{fig:Nanowires}(c) the ratio $\rho_a/\rho_b$ of the resistivities measured at 300 K on untwinned, underdoped ($p = 0.125$) nanowires are shown for two values of the film thickness $t$. Nanowires with $t=50$ nm are characterized by a $\rho$ anisotropy which is in perfect agreement with that expected in untwinned single crystals, at the same oxygen level \cite{ando2002electrical}. The resistivity ratio doubles in nanowires with $t=10$ nm. Here, a nematicity is present, in addition to the trivial anisotropy due to the chains. The origin of this nematicity can possibly be related to the strain, which is much more effective in ultrathin films, where interface effects play a crucial role (as previously highlighted in Sec. \ref{sec: strain}).

\section{Summary and Conclusions}

To conclude, by using a proper substrate preparation, we have succeeded in the growth of \textit{untwinned}  and \textit{compressively strained}  YBCO thin films and nanostructures. 

YBCO is expected to grow twinned on (110) oriented MgO substrates. To modify the twinning state, we performed a thermal treatment in oxygen atmosphere of the substrates, to induce the reconstruction of the (110) surface, which is known to be unstable. This leads to the formation of elongated facets of shallow, high Miller index planes. When a YBCO film is deposited on top of these annealed substrates, the facets act \textcolor{black}{in the same way as} the step edges of vicinal angle (001) oriented STO substrates, i.e. they favor the alignment of the CuO chains along the elongated side of the facets. 

A peculiar strain occurs at the film-substrate interface, as a consequence of the large mismatch between the in-plane lattice parameters. \textcolor{black}{Along the $b$ direction, a variation of the in-plane bond lengths occurs: the CuO chains are compressively strained, and present a buckling with respect to the substrate surface, which is commonly associated to a distortion of the oxygen octahedra.}

Films with different oxygen doping from the strongly underdoped up to the slightly overdoped level  have been patterned into nanowires, with dimensions down to 50 nm. Our preliminary results show that the untwinning state of YBCO is preserved down to the nanoscale. Indeed, we have focused on the values of the critical current density and \textcolor{black}{of the room-temperature resistivity} of nanowires patterned either along $a$ or along $b$. The intrinsic anisotropy of the transport properties, deriving by the presence of the CuO chains along the $b$ axis direction, remains, with values in very good agreement with those of untwinned single crystals \cite{basov1995plane, ando2002electrical}. Moreover, we got some hints of the presence of additional, unexpected, anisotropies, induced by the strain, in ultrathin films where the role of the substrates in modifying the length and shape of the relaxed YBCO unit cell is more relevant.

The latter result shows that, when strain effects become dominant, the YBCO properties are modified with respect to the relaxed, bulk form. This occurrence is of high relevance. In our 50 nm thick films, where the role of the strain is marginal, we have recently observed bulk-like charge order \cite{arpaia2018dynamical}.  The compression, by pressure or mechanical stress, of untwinned single crystals has been  shown to strongly modify - either suppressing or enhancing - several intertwining orders, such as the charge order, in the underdoped region of the cuprate phase diagram. Our films and nanostructures can therefore represent a clean system where one can tune  these orders (CDW, nematicity, etc.) down to the nanoscale by strain.

\begin{acknowledgements}
This work has been supported by the Swedish Research Council (VR) and by the Knut and Alice Wallenberg Foundation. R. A. is supported by the Swedish Research Council (VR) under the project 2017-00382. R.A. is grateful to Dr Carmela Aruta for fruitful discussions at the early stage of this work, and to Henrik Frederiksen for technical support. The DFT calculations were in part performed on resources at Chalmers Centre for Computational Science and Engineering (C3SE) provided by the Swedish National Infrastructure for Computing (SNIC).
\end{acknowledgements}
\bibliography{biblio}

\begin{thebibliography}{77}%
\makeatletter
\providecommand \@ifxundefined [1]{%
 \@ifx{#1\undefined}
}%
\providecommand \@ifnum [1]{%
 \ifnum #1\expandafter \@firstoftwo
 \else \expandafter \@secondoftwo
 \fi
}%
\providecommand \@ifx [1]{%
 \ifx #1\expandafter \@firstoftwo
 \else \expandafter \@secondoftwo
 \fi
}%
\providecommand \natexlab [1]{#1}%
\providecommand \enquote  [1]{``#1''}%
\providecommand \bibnamefont  [1]{#1}%
\providecommand \bibfnamefont [1]{#1}%
\providecommand \citenamefont [1]{#1}%
\providecommand \href@noop [0]{\@secondoftwo}%
\providecommand \href [0]{\begingroup \@sanitize@url \@href}%
\providecommand \@href[1]{\@@startlink{#1}\@@href}%
\providecommand \@@href[1]{\endgroup#1\@@endlink}%
\providecommand \@sanitize@url [0]{\catcode `\\12\catcode `\$12\catcode
  `\&12\catcode `\#12\catcode `\^12\catcode `\_12\catcode `\%12\relax}%
\providecommand \@@startlink[1]{}%
\providecommand \@@endlink[0]{}%
\providecommand \url  [0]{\begingroup\@sanitize@url \@url }%
\providecommand \@url [1]{\endgroup\@href {#1}{\urlprefix }}%
\providecommand \urlprefix  [0]{URL }%
\providecommand \Eprint [0]{\href }%
\providecommand \doibase [0]{http://dx.doi.org/}%
\providecommand \selectlanguage [0]{\@gobble}%
\providecommand \bibinfo  [0]{\@secondoftwo}%
\providecommand \bibfield  [0]{\@secondoftwo}%
\providecommand \translation [1]{[#1]}%
\providecommand \BibitemOpen [0]{}%
\providecommand \bibitemStop [0]{}%
\providecommand \bibitemNoStop [0]{.\EOS\space}%
\providecommand \EOS [0]{\spacefactor3000\relax}%
\providecommand \BibitemShut  [1]{\csname bibitem#1\endcsname}%
\let\auto@bib@innerbib\@empty
\bibitem [{\citenamefont {Keimer}\ \emph {et~al.}(2015)\citenamefont {Keimer},
  \citenamefont {Kivelson}, \citenamefont {Norman}, \citenamefont {Uchida},\
  and\ \citenamefont {Zaanen}}]{keimer2015quantum}%
  \BibitemOpen
  \bibfield  {author} {\bibinfo {author} {\bibfnamefont {B.}~\bibnamefont
  {Keimer}}, \bibinfo {author} {\bibfnamefont {S.~A.}\ \bibnamefont
  {Kivelson}}, \bibinfo {author} {\bibfnamefont {M.~R.}\ \bibnamefont
  {Norman}}, \bibinfo {author} {\bibfnamefont {S.}~\bibnamefont {Uchida}}, \
  and\ \bibinfo {author} {\bibfnamefont {J.}~\bibnamefont {Zaanen}},\ }\href
  {\doibase 10.1038/nature14165} {\bibfield  {journal} {\bibinfo  {journal}
  {Nature}\ }\textbf {\bibinfo {volume} {518}},\ \bibinfo {pages} {179}
  (\bibinfo {year} {2015})}\BibitemShut {NoStop}%
\bibitem [{\citenamefont {Hardy}\ \emph {et~al.}(2010)\citenamefont {Hardy},
  \citenamefont {Hillier}, \citenamefont {Meingast}, \citenamefont {Colson},
  \citenamefont {Li}, \citenamefont {Bari{\v{s}}i{\'c}}, \citenamefont {Yu},
  \citenamefont {Zhao}, \citenamefont {Greven},\ and\ \citenamefont
  {Schilling}}]{hardy2010enhancement}%
  \BibitemOpen
  \bibfield  {author} {\bibinfo {author} {\bibfnamefont {F.}~\bibnamefont
  {Hardy}}, \bibinfo {author} {\bibfnamefont {N.}~\bibnamefont {Hillier}},
  \bibinfo {author} {\bibfnamefont {C.}~\bibnamefont {Meingast}}, \bibinfo
  {author} {\bibfnamefont {D.}~\bibnamefont {Colson}}, \bibinfo {author}
  {\bibfnamefont {Y.}~\bibnamefont {Li}}, \bibinfo {author} {\bibfnamefont
  {N.}~\bibnamefont {Bari{\v{s}}i{\'c}}}, \bibinfo {author} {\bibfnamefont
  {G.}~\bibnamefont {Yu}}, \bibinfo {author} {\bibfnamefont {X.}~\bibnamefont
  {Zhao}}, \bibinfo {author} {\bibfnamefont {M.}~\bibnamefont {Greven}}, \ and\
  \bibinfo {author} {\bibfnamefont {J.}~\bibnamefont {Schilling}},\ }\href
  {\doibase 10.1103/PhysRevLett.105.167002} {\bibfield  {journal} {\bibinfo
  {journal} {Phys. Rev. Lett.}\ }\textbf {\bibinfo {volume} {105}},\ \bibinfo
  {pages} {167002} (\bibinfo {year} {2010})}\BibitemShut {NoStop}%
\bibitem [{\citenamefont {Wu}\ \emph {et~al.}(2017)\citenamefont {Wu},
  \citenamefont {Bollinger}, \citenamefont {He},\ and\ \citenamefont
  {Bo{\v{z}}ovi{\'c}}}]{wu2017spontaneous}%
  \BibitemOpen
  \bibfield  {author} {\bibinfo {author} {\bibfnamefont {J.}~\bibnamefont
  {Wu}}, \bibinfo {author} {\bibfnamefont {A.}~\bibnamefont {Bollinger}},
  \bibinfo {author} {\bibfnamefont {X.}~\bibnamefont {He}}, \ and\ \bibinfo
  {author} {\bibfnamefont {I.}~\bibnamefont {Bo{\v{z}}ovi{\'c}}},\ }\href
  {\doibase 10.1038/nature23290} {\bibfield  {journal} {\bibinfo  {journal}
  {Nature}\ }\textbf {\bibinfo {volume} {547}},\ \bibinfo {pages} {432}
  (\bibinfo {year} {2017})}\BibitemShut {NoStop}%
\bibitem [{\citenamefont {Ivashko}\ \emph {et~al.}(2019)\citenamefont
  {Ivashko}, \citenamefont {Horio}, \citenamefont {Wan}, \citenamefont
  {Christensen}, \citenamefont {McNally}, \citenamefont {Paris}, \citenamefont
  {Tseng}, \citenamefont {Shaik}, \citenamefont {R{\o}nnow}, \citenamefont
  {Wei}, \citenamefont {Adamo}, \citenamefont {Lichtensteiger}, \citenamefont
  {Gibert}, \citenamefont {Beasley}, \citenamefont {Shen}, \citenamefont
  {Tomczak}, \citenamefont {Schmitt},\ and\ \citenamefont
  {Chang}}]{ivashko2019strain}%
  \BibitemOpen
  \bibfield  {author} {\bibinfo {author} {\bibfnamefont {O.}~\bibnamefont
  {Ivashko}}, \bibinfo {author} {\bibfnamefont {M.}~\bibnamefont {Horio}},
  \bibinfo {author} {\bibfnamefont {W.}~\bibnamefont {Wan}}, \bibinfo {author}
  {\bibfnamefont {N.}~\bibnamefont {Christensen}}, \bibinfo {author}
  {\bibfnamefont {D.}~\bibnamefont {McNally}}, \bibinfo {author} {\bibfnamefont
  {E.}~\bibnamefont {Paris}}, \bibinfo {author} {\bibfnamefont
  {Y.}~\bibnamefont {Tseng}}, \bibinfo {author} {\bibfnamefont
  {N.}~\bibnamefont {Shaik}}, \bibinfo {author} {\bibfnamefont
  {H.}~\bibnamefont {R{\o}nnow}}, \bibinfo {author} {\bibfnamefont
  {H.}~\bibnamefont {Wei}}, \bibinfo {author} {\bibfnamefont {C.}~\bibnamefont
  {Adamo}}, \bibinfo {author} {\bibfnamefont {C.}~\bibnamefont
  {Lichtensteiger}}, \bibinfo {author} {\bibfnamefont {M.}~\bibnamefont
  {Gibert}}, \bibinfo {author} {\bibfnamefont {M.}~\bibnamefont {Beasley}},
  \bibinfo {author} {\bibfnamefont {K.}~\bibnamefont {Shen}}, \bibinfo {author}
  {\bibfnamefont {J.}~\bibnamefont {Tomczak}}, \bibinfo {author} {\bibfnamefont
  {T.}~\bibnamefont {Schmitt}}, \ and\ \bibinfo {author} {\bibfnamefont
  {J.}~\bibnamefont {Chang}},\ }\href {\doibase 10.1038/s41467-019-08664-6}
  {\bibfield  {journal} {\bibinfo  {journal} {Nat. Commun.}\ }\textbf {\bibinfo
  {volume} {10}},\ \bibinfo {pages} {786} (\bibinfo {year} {2019})}\BibitemShut
  {NoStop}%
\bibitem [{\citenamefont {Bluschke}\ \emph {et~al.}(2018)\citenamefont
  {Bluschke}, \citenamefont {Frano}, \citenamefont {Schierle}, \citenamefont
  {Putzky}, \citenamefont {Ghorbani}, \citenamefont {Ortiz}, \citenamefont
  {Suzuki}, \citenamefont {Christiani}, \citenamefont {Logvenov}, \citenamefont
  {Weschke}, \citenamefont {Birgeneau}, \citenamefont {da~Silva~Neto},
  \citenamefont {Minola}, \citenamefont {Blanco-Canosa},\ and\ \citenamefont
  {Keimer}}]{bluschke2018stabilization}%
  \BibitemOpen
  \bibfield  {author} {\bibinfo {author} {\bibfnamefont {M.}~\bibnamefont
  {Bluschke}}, \bibinfo {author} {\bibfnamefont {A.}~\bibnamefont {Frano}},
  \bibinfo {author} {\bibfnamefont {E.}~\bibnamefont {Schierle}}, \bibinfo
  {author} {\bibfnamefont {D.}~\bibnamefont {Putzky}}, \bibinfo {author}
  {\bibfnamefont {F.}~\bibnamefont {Ghorbani}}, \bibinfo {author}
  {\bibfnamefont {R.}~\bibnamefont {Ortiz}}, \bibinfo {author} {\bibfnamefont
  {H.}~\bibnamefont {Suzuki}}, \bibinfo {author} {\bibfnamefont
  {G.}~\bibnamefont {Christiani}}, \bibinfo {author} {\bibfnamefont
  {G.}~\bibnamefont {Logvenov}}, \bibinfo {author} {\bibfnamefont
  {E.}~\bibnamefont {Weschke}}, \bibinfo {author} {\bibfnamefont
  {R.}~\bibnamefont {Birgeneau}}, \bibinfo {author} {\bibfnamefont
  {E.}~\bibnamefont {da~Silva~Neto}}, \bibinfo {author} {\bibfnamefont
  {M.}~\bibnamefont {Minola}}, \bibinfo {author} {\bibfnamefont
  {S.}~\bibnamefont {Blanco-Canosa}}, \ and\ \bibinfo {author} {\bibfnamefont
  {B.}~\bibnamefont {Keimer}},\ }\href {\doibase 10.1038/s41467-018-05434-8}
  {\bibfield  {journal} {\bibinfo  {journal} {Nat. Commun.}\ }\textbf {\bibinfo
  {volume} {9}},\ \bibinfo {pages} {2978} (\bibinfo {year} {2018})}\BibitemShut
  {NoStop}%
\bibitem [{\citenamefont {Huang}\ \emph {et~al.}(2018)\citenamefont {Huang},
  \citenamefont {Jang}, \citenamefont {Fujita}, \citenamefont {Nishizaki},
  \citenamefont {Lin}, \citenamefont {Wang}, \citenamefont {Ying},
  \citenamefont {Smith}, \citenamefont {Kenney-Benson}, \citenamefont {Shen},
  \citenamefont {Mao}, \citenamefont {Kao}, \citenamefont {Liu},\ and\
  \citenamefont {Lee}}]{huang2018modification}%
  \BibitemOpen
  \bibfield  {author} {\bibinfo {author} {\bibfnamefont {H.}~\bibnamefont
  {Huang}}, \bibinfo {author} {\bibfnamefont {H.}~\bibnamefont {Jang}},
  \bibinfo {author} {\bibfnamefont {M.}~\bibnamefont {Fujita}}, \bibinfo
  {author} {\bibfnamefont {T.}~\bibnamefont {Nishizaki}}, \bibinfo {author}
  {\bibfnamefont {Y.}~\bibnamefont {Lin}}, \bibinfo {author} {\bibfnamefont
  {J.}~\bibnamefont {Wang}}, \bibinfo {author} {\bibfnamefont {J.}~\bibnamefont
  {Ying}}, \bibinfo {author} {\bibfnamefont {J.}~\bibnamefont {Smith}},
  \bibinfo {author} {\bibfnamefont {C.}~\bibnamefont {Kenney-Benson}}, \bibinfo
  {author} {\bibfnamefont {G.}~\bibnamefont {Shen}}, \bibinfo {author}
  {\bibfnamefont {W.}~\bibnamefont {Mao}}, \bibinfo {author} {\bibfnamefont
  {C.-C.}\ \bibnamefont {Kao}}, \bibinfo {author} {\bibfnamefont {Y.-J.}\
  \bibnamefont {Liu}}, \ and\ \bibinfo {author} {\bibfnamefont {J.-S.}\
  \bibnamefont {Lee}},\ }\href {\doibase 10.1103/PhysRevB.97.174508} {\bibfield
   {journal} {\bibinfo  {journal} {Phys. Rev. B}\ }\textbf {\bibinfo {volume}
  {97}},\ \bibinfo {pages} {174508} (\bibinfo {year} {2018})}\BibitemShut
  {NoStop}%
\bibitem [{\citenamefont {Kim}\ \emph {et~al.}(2018)\citenamefont {Kim},
  \citenamefont {Souliou}, \citenamefont {Barber}, \citenamefont
  {Lefran{\c{c}}ois}, \citenamefont {Minola}, \citenamefont {Tortora},
  \citenamefont {Heid}, \citenamefont {Nandi}, \citenamefont {Borzi},
  \citenamefont {Garbarino}, \citenamefont {Bosak}, \citenamefont {Porras},
  \citenamefont {Loew}, \citenamefont {K{\"o}nig}, \citenamefont {Moll},
  \citenamefont {Mackenzie}, \citenamefont {Keimer}, \citenamefont {Hicks},\
  and\ \citenamefont {Le~Tacon}}]{kim2018uniaxial}%
  \BibitemOpen
  \bibfield  {author} {\bibinfo {author} {\bibfnamefont {H.-H.}\ \bibnamefont
  {Kim}}, \bibinfo {author} {\bibfnamefont {S.}~\bibnamefont {Souliou}},
  \bibinfo {author} {\bibfnamefont {M.}~\bibnamefont {Barber}}, \bibinfo
  {author} {\bibfnamefont {E.}~\bibnamefont {Lefran{\c{c}}ois}}, \bibinfo
  {author} {\bibfnamefont {M.}~\bibnamefont {Minola}}, \bibinfo {author}
  {\bibfnamefont {M.}~\bibnamefont {Tortora}}, \bibinfo {author} {\bibfnamefont
  {R.}~\bibnamefont {Heid}}, \bibinfo {author} {\bibfnamefont {N.}~\bibnamefont
  {Nandi}}, \bibinfo {author} {\bibfnamefont {R.}~\bibnamefont {Borzi}},
  \bibinfo {author} {\bibfnamefont {G.}~\bibnamefont {Garbarino}}, \bibinfo
  {author} {\bibfnamefont {A.}~\bibnamefont {Bosak}}, \bibinfo {author}
  {\bibfnamefont {J.}~\bibnamefont {Porras}}, \bibinfo {author} {\bibfnamefont
  {T.}~\bibnamefont {Loew}}, \bibinfo {author} {\bibfnamefont {M.}~\bibnamefont
  {K{\"o}nig}}, \bibinfo {author} {\bibfnamefont {P.}~\bibnamefont {Moll}},
  \bibinfo {author} {\bibfnamefont {A.}~\bibnamefont {Mackenzie}}, \bibinfo
  {author} {\bibfnamefont {B.}~\bibnamefont {Keimer}}, \bibinfo {author}
  {\bibfnamefont {C.}~\bibnamefont {Hicks}}, \ and\ \bibinfo {author}
  {\bibfnamefont {M.}~\bibnamefont {Le~Tacon}},\ }\href {\doibase
  10.1126/science.aat4708} {\bibfield  {journal} {\bibinfo  {journal}
  {Science}\ }\textbf {\bibinfo {volume} {362}},\ \bibinfo {pages} {1040}
  (\bibinfo {year} {2018})}\BibitemShut {NoStop}%
\bibitem [{\citenamefont {Ghiringhelli}\ \emph {et~al.}(2012)\citenamefont
  {Ghiringhelli}, \citenamefont {Le~Tacon}, \citenamefont {Minola},
  \citenamefont {Blanco-Canosa}, \citenamefont {Mazzoli}, \citenamefont
  {Brookes}, \citenamefont {De~Luca}, \citenamefont {Frano}, \citenamefont
  {Hawthorn}, \citenamefont {He}, \citenamefont {Loew}, \citenamefont
  {Moretti~Sala}, \citenamefont {Peets}, \citenamefont {Salluzzo},
  \citenamefont {Schierle}, \citenamefont {Sutarto}, \citenamefont {Sawatzky},
  \citenamefont {Weschke}, \citenamefont {Keimer},\ and\ \citenamefont
  {Braicovich}}]{ghiringhelli2012long}%
  \BibitemOpen
  \bibfield  {author} {\bibinfo {author} {\bibfnamefont {G.}~\bibnamefont
  {Ghiringhelli}}, \bibinfo {author} {\bibfnamefont {M.}~\bibnamefont
  {Le~Tacon}}, \bibinfo {author} {\bibfnamefont {M.}~\bibnamefont {Minola}},
  \bibinfo {author} {\bibfnamefont {S.}~\bibnamefont {Blanco-Canosa}}, \bibinfo
  {author} {\bibfnamefont {C.}~\bibnamefont {Mazzoli}}, \bibinfo {author}
  {\bibfnamefont {N.}~\bibnamefont {Brookes}}, \bibinfo {author} {\bibfnamefont
  {G.}~\bibnamefont {De~Luca}}, \bibinfo {author} {\bibfnamefont
  {A.}~\bibnamefont {Frano}}, \bibinfo {author} {\bibfnamefont
  {D.}~\bibnamefont {Hawthorn}}, \bibinfo {author} {\bibfnamefont
  {F.}~\bibnamefont {He}}, \bibinfo {author} {\bibfnamefont {T.}~\bibnamefont
  {Loew}}, \bibinfo {author} {\bibfnamefont {M.}~\bibnamefont {Moretti~Sala}},
  \bibinfo {author} {\bibfnamefont {D.}~\bibnamefont {Peets}}, \bibinfo
  {author} {\bibfnamefont {M.}~\bibnamefont {Salluzzo}}, \bibinfo {author}
  {\bibfnamefont {E.}~\bibnamefont {Schierle}}, \bibinfo {author}
  {\bibfnamefont {R.}~\bibnamefont {Sutarto}}, \bibinfo {author} {\bibfnamefont
  {G.}~\bibnamefont {Sawatzky}}, \bibinfo {author} {\bibfnamefont
  {E.}~\bibnamefont {Weschke}}, \bibinfo {author} {\bibfnamefont
  {B.}~\bibnamefont {Keimer}}, \ and\ \bibinfo {author} {\bibfnamefont
  {L.}~\bibnamefont {Braicovich}},\ }\href {\doibase 10.1126/science.1223532}
  {\bibfield  {journal} {\bibinfo  {journal} {Science}\ }\textbf {\bibinfo
  {volume} {337}},\ \bibinfo {pages} {821} (\bibinfo {year}
  {2012})}\BibitemShut {NoStop}%
\bibitem [{\citenamefont {Chang}\ \emph {et~al.}(2012)\citenamefont {Chang},
  \citenamefont {Blackburn}, \citenamefont {Holmes}, \citenamefont
  {Christensen}, \citenamefont {Larsen}, \citenamefont {Mesot}, \citenamefont
  {Liang}, \citenamefont {Bonn}, \citenamefont {Hardy}, \citenamefont
  {Watenphul}, \citenamefont {Zimmermann}, \citenamefont {Forgan},\ and\
  \citenamefont {Hayden}}]{chang2012direct}%
  \BibitemOpen
  \bibfield  {author} {\bibinfo {author} {\bibfnamefont {J.}~\bibnamefont
  {Chang}}, \bibinfo {author} {\bibfnamefont {E.}~\bibnamefont {Blackburn}},
  \bibinfo {author} {\bibfnamefont {A.}~\bibnamefont {Holmes}}, \bibinfo
  {author} {\bibfnamefont {N.~B.}\ \bibnamefont {Christensen}}, \bibinfo
  {author} {\bibfnamefont {J.}~\bibnamefont {Larsen}}, \bibinfo {author}
  {\bibfnamefont {J.}~\bibnamefont {Mesot}}, \bibinfo {author} {\bibfnamefont
  {R.}~\bibnamefont {Liang}}, \bibinfo {author} {\bibfnamefont
  {D.}~\bibnamefont {Bonn}}, \bibinfo {author} {\bibfnamefont {W.}~\bibnamefont
  {Hardy}}, \bibinfo {author} {\bibfnamefont {A.}~\bibnamefont {Watenphul}},
  \bibinfo {author} {\bibfnamefont {M.}~\bibnamefont {Zimmermann}}, \bibinfo
  {author} {\bibfnamefont {E.}~\bibnamefont {Forgan}}, \ and\ \bibinfo {author}
  {\bibfnamefont {S.}~\bibnamefont {Hayden}},\ }\href {\doibase
  10.1038/nphys2456} {\bibfield  {journal} {\bibinfo  {journal} {Nat. Phys.}\
  }\textbf {\bibinfo {volume} {8}},\ \bibinfo {pages} {871} (\bibinfo {year}
  {2012})}\BibitemShut {NoStop}%
\bibitem [{\citenamefont {Gerber}\ \emph {et~al.}(2015)\citenamefont {Gerber},
  \citenamefont {Jang}, \citenamefont {Nojiri}, \citenamefont {Matsuzawa},
  \citenamefont {Yasumura}, \citenamefont {Bonn}, \citenamefont {Liang},
  \citenamefont {Hardy}, \citenamefont {Islam}, \citenamefont {Mehta},
  \citenamefont {Song}, \citenamefont {Sikorski}, \citenamefont {Stefanescu},
  \citenamefont {Feng}, \citenamefont {Kivelson}, \citenamefont {Devereaux},
  \citenamefont {Shen}, \citenamefont {Kao}, \citenamefont {Lee}, \citenamefont
  {Zhu},\ and\ \citenamefont {Lee}}]{gerber2015three}%
  \BibitemOpen
  \bibfield  {author} {\bibinfo {author} {\bibfnamefont {S.}~\bibnamefont
  {Gerber}}, \bibinfo {author} {\bibfnamefont {H.}~\bibnamefont {Jang}},
  \bibinfo {author} {\bibfnamefont {H.}~\bibnamefont {Nojiri}}, \bibinfo
  {author} {\bibfnamefont {S.}~\bibnamefont {Matsuzawa}}, \bibinfo {author}
  {\bibfnamefont {H.}~\bibnamefont {Yasumura}}, \bibinfo {author}
  {\bibfnamefont {D.}~\bibnamefont {Bonn}}, \bibinfo {author} {\bibfnamefont
  {R.}~\bibnamefont {Liang}}, \bibinfo {author} {\bibfnamefont
  {W.}~\bibnamefont {Hardy}}, \bibinfo {author} {\bibfnamefont
  {Z.}~\bibnamefont {Islam}}, \bibinfo {author} {\bibfnamefont
  {A.}~\bibnamefont {Mehta}}, \bibinfo {author} {\bibfnamefont
  {S.}~\bibnamefont {Song}}, \bibinfo {author} {\bibfnamefont {M.}~\bibnamefont
  {Sikorski}}, \bibinfo {author} {\bibfnamefont {D.}~\bibnamefont
  {Stefanescu}}, \bibinfo {author} {\bibfnamefont {Y.}~\bibnamefont {Feng}},
  \bibinfo {author} {\bibfnamefont {S.}~\bibnamefont {Kivelson}}, \bibinfo
  {author} {\bibfnamefont {T.}~\bibnamefont {Devereaux}}, \bibinfo {author}
  {\bibfnamefont {Z.-X.}\ \bibnamefont {Shen}}, \bibinfo {author}
  {\bibfnamefont {C.-C.}\ \bibnamefont {Kao}}, \bibinfo {author} {\bibfnamefont
  {W.-S.}\ \bibnamefont {Lee}}, \bibinfo {author} {\bibfnamefont
  {D.}~\bibnamefont {Zhu}}, \ and\ \bibinfo {author} {\bibfnamefont {J.-S.}\
  \bibnamefont {Lee}},\ }\href {\doibase 10.1126/science.aac6257} {\bibfield
  {journal} {\bibinfo  {journal} {Science}\ }\textbf {\bibinfo {volume}
  {350}},\ \bibinfo {pages} {949} (\bibinfo {year} {2015})}\BibitemShut
  {NoStop}%
\bibitem [{\citenamefont {Comin}\ and\ \citenamefont
  {Damascelli}(2016)}]{comin2016resonant}%
  \BibitemOpen
  \bibfield  {author} {\bibinfo {author} {\bibfnamefont {R.}~\bibnamefont
  {Comin}}\ and\ \bibinfo {author} {\bibfnamefont {A.}~\bibnamefont
  {Damascelli}},\ }\href {\doibase 10.1146/annurev-conmatphys-031115-011401}
  {\bibfield  {journal} {\bibinfo  {journal} {Annu. Rev. Condens. Matter
  Phys.}\ }\textbf {\bibinfo {volume} {7}},\ \bibinfo {pages} {369} (\bibinfo
  {year} {2016})}\BibitemShut {NoStop}%
\bibitem [{\citenamefont {Loret}\ \emph {et~al.}(2019)\citenamefont {Loret},
  \citenamefont {Auvray}, \citenamefont {Gallais}, \citenamefont {Cazayous},
  \citenamefont {Forget}, \citenamefont {Colson}, \citenamefont {Julien},
  \citenamefont {Paul}, \citenamefont {Civelli},\ and\ \citenamefont
  {Sacuto}}]{loret2019intimate}%
  \BibitemOpen
  \bibfield  {author} {\bibinfo {author} {\bibfnamefont {B.}~\bibnamefont
  {Loret}}, \bibinfo {author} {\bibfnamefont {N.}~\bibnamefont {Auvray}},
  \bibinfo {author} {\bibfnamefont {Y.}~\bibnamefont {Gallais}}, \bibinfo
  {author} {\bibfnamefont {M.}~\bibnamefont {Cazayous}}, \bibinfo {author}
  {\bibfnamefont {A.}~\bibnamefont {Forget}}, \bibinfo {author} {\bibfnamefont
  {D.}~\bibnamefont {Colson}}, \bibinfo {author} {\bibfnamefont {M.-H.}\
  \bibnamefont {Julien}}, \bibinfo {author} {\bibfnamefont {I.}~\bibnamefont
  {Paul}}, \bibinfo {author} {\bibfnamefont {M.}~\bibnamefont {Civelli}}, \
  and\ \bibinfo {author} {\bibfnamefont {A.}~\bibnamefont {Sacuto}},\
  }\href@noop {} {\bibfield  {journal} {\bibinfo  {journal} {Nat. Phys.}\
  }\textbf {\bibinfo {volume} {15}},\ \bibinfo {pages} {771} (\bibinfo {year}
  {2019})}\BibitemShut {NoStop}%
\bibitem [{\citenamefont {Sadewasser}\ \emph {et~al.}(2000)\citenamefont
  {Sadewasser}, \citenamefont {Schilling}, \citenamefont {Paulikas},\ and\
  \citenamefont {Veal}}]{sadewasser2000pressure}%
  \BibitemOpen
  \bibfield  {author} {\bibinfo {author} {\bibfnamefont {S.}~\bibnamefont
  {Sadewasser}}, \bibinfo {author} {\bibfnamefont {J.}~\bibnamefont
  {Schilling}}, \bibinfo {author} {\bibfnamefont {A.}~\bibnamefont {Paulikas}},
  \ and\ \bibinfo {author} {\bibfnamefont {B.}~\bibnamefont {Veal}},\ }\href
  {\doibase 10.1103/PhysRevB.61.741} {\bibfield  {journal} {\bibinfo  {journal}
  {Phys. Rev. B}\ }\textbf {\bibinfo {volume} {61}},\ \bibinfo {pages} {741}
  (\bibinfo {year} {2000})}\BibitemShut {NoStop}%
\bibitem [{\citenamefont {Cyr-Choini{\`e}re}\ \emph {et~al.}(2018)\citenamefont
  {Cyr-Choini{\`e}re}, \citenamefont {LeBoeuf}, \citenamefont {Badoux},
  \citenamefont {Dufour-Beaus{\'e}jour}, \citenamefont {Bonn}, \citenamefont
  {Hardy}, \citenamefont {Liang}, \citenamefont {Graf}, \citenamefont
  {Doiron-Leyraud},\ and\ \citenamefont {Taillefer}}]{cyr2018sensitivity}%
  \BibitemOpen
  \bibfield  {author} {\bibinfo {author} {\bibfnamefont {O.}~\bibnamefont
  {Cyr-Choini{\`e}re}}, \bibinfo {author} {\bibfnamefont {D.}~\bibnamefont
  {LeBoeuf}}, \bibinfo {author} {\bibfnamefont {S.}~\bibnamefont {Badoux}},
  \bibinfo {author} {\bibfnamefont {S.}~\bibnamefont {Dufour-Beaus{\'e}jour}},
  \bibinfo {author} {\bibfnamefont {D.}~\bibnamefont {Bonn}}, \bibinfo {author}
  {\bibfnamefont {W.}~\bibnamefont {Hardy}}, \bibinfo {author} {\bibfnamefont
  {R.}~\bibnamefont {Liang}}, \bibinfo {author} {\bibfnamefont
  {D.}~\bibnamefont {Graf}}, \bibinfo {author} {\bibfnamefont {N.}~\bibnamefont
  {Doiron-Leyraud}}, \ and\ \bibinfo {author} {\bibfnamefont {L.}~\bibnamefont
  {Taillefer}},\ }\href {\doibase 10.1103/PhysRevB.98.064513} {\bibfield
  {journal} {\bibinfo  {journal} {Phys. Rev. B}\ }\textbf {\bibinfo {volume}
  {98}},\ \bibinfo {pages} {064513} (\bibinfo {year} {2018})}\BibitemShut
  {NoStop}%
\bibitem [{\citenamefont {Steinborn}\ \emph {et~al.}(1994)\citenamefont
  {Steinborn}, \citenamefont {Miehe}, \citenamefont {Wiesner}, \citenamefont
  {Brecht}, \citenamefont {Fuess}, \citenamefont {Wirth}, \citenamefont
  {Schulte}, \citenamefont {Speckmann}, \citenamefont {Adrian}, \citenamefont
  {Maul}, \citenamefont {Petersen}, \citenamefont {Blau},\ and\ \citenamefont
  {McConnel}}]{steinborn1994twinning}%
  \BibitemOpen
  \bibfield  {author} {\bibinfo {author} {\bibfnamefont {T.}~\bibnamefont
  {Steinborn}}, \bibinfo {author} {\bibfnamefont {G.}~\bibnamefont {Miehe}},
  \bibinfo {author} {\bibfnamefont {J.}~\bibnamefont {Wiesner}}, \bibinfo
  {author} {\bibfnamefont {E.}~\bibnamefont {Brecht}}, \bibinfo {author}
  {\bibfnamefont {H.}~\bibnamefont {Fuess}}, \bibinfo {author} {\bibfnamefont
  {G.}~\bibnamefont {Wirth}}, \bibinfo {author} {\bibfnamefont
  {B.}~\bibnamefont {Schulte}}, \bibinfo {author} {\bibfnamefont
  {M.}~\bibnamefont {Speckmann}}, \bibinfo {author} {\bibfnamefont
  {H.}~\bibnamefont {Adrian}}, \bibinfo {author} {\bibfnamefont
  {M.}~\bibnamefont {Maul}}, \bibinfo {author} {\bibfnamefont {K.}~\bibnamefont
  {Petersen}}, \bibinfo {author} {\bibfnamefont {W.}~\bibnamefont {Blau}}, \
  and\ \bibinfo {author} {\bibfnamefont {M.}~\bibnamefont {McConnel}},\ }\href
  {\doibase 10.1016/0921-4534(94)90906-7} {\bibfield  {journal} {\bibinfo
  {journal} {Physica C}\ }\textbf {\bibinfo {volume} {220}},\ \bibinfo {pages}
  {219} (\bibinfo {year} {1994})}\BibitemShut {NoStop}%
\bibitem [{\citenamefont {Schweitzer}\ \emph {et~al.}(1996)\citenamefont
  {Schweitzer}, \citenamefont {Bollmeier}, \citenamefont {Stritzker},\ and\
  \citenamefont {Rauschenbach}}]{schweitzer1996twinning}%
  \BibitemOpen
  \bibfield  {author} {\bibinfo {author} {\bibfnamefont {D.}~\bibnamefont
  {Schweitzer}}, \bibinfo {author} {\bibfnamefont {T.}~\bibnamefont
  {Bollmeier}}, \bibinfo {author} {\bibfnamefont {B.}~\bibnamefont
  {Stritzker}}, \ and\ \bibinfo {author} {\bibfnamefont {B.}~\bibnamefont
  {Rauschenbach}},\ }\href {\doibase 10.1016/0040-6090(95)08210-7} {\bibfield
  {journal} {\bibinfo  {journal} {Thin Solid Films}\ }\textbf {\bibinfo
  {volume} {280}},\ \bibinfo {pages} {147} (\bibinfo {year}
  {1996})}\BibitemShut {NoStop}%
\bibitem [{\citenamefont {Bernstein}\ and\ \citenamefont
  {Hamet}(2004)}]{bernstein2004vortex}%
  \BibitemOpen
  \bibfield  {author} {\bibinfo {author} {\bibfnamefont {P.}~\bibnamefont
  {Bernstein}}\ and\ \bibinfo {author} {\bibfnamefont {J.}~\bibnamefont
  {Hamet}},\ }\href {\doibase 10.1063/1.1642761} {\bibfield  {journal}
  {\bibinfo  {journal} {J. Appl. Phys.}\ }\textbf {\bibinfo {volume} {95}},\
  \bibinfo {pages} {2569} (\bibinfo {year} {2004})}\BibitemShut {NoStop}%
\bibitem [{\citenamefont {Frano}\ \emph {et~al.}(2016)\citenamefont {Frano},
  \citenamefont {Blanco-Canosa}, \citenamefont {Schierle}, \citenamefont {Lu},
  \citenamefont {Wu}, \citenamefont {Bluschke}, \citenamefont {Minola},
  \citenamefont {Christiani}, \citenamefont {Habermeier}, \citenamefont
  {Logvenov} \emph {et~al.}}]{frano2016long}%
  \BibitemOpen
  \bibfield  {author} {\bibinfo {author} {\bibfnamefont {A.}~\bibnamefont
  {Frano}}, \bibinfo {author} {\bibfnamefont {S.}~\bibnamefont
  {Blanco-Canosa}}, \bibinfo {author} {\bibfnamefont {E.}~\bibnamefont
  {Schierle}}, \bibinfo {author} {\bibfnamefont {Y.}~\bibnamefont {Lu}},
  \bibinfo {author} {\bibfnamefont {M.}~\bibnamefont {Wu}}, \bibinfo {author}
  {\bibfnamefont {M.}~\bibnamefont {Bluschke}}, \bibinfo {author}
  {\bibfnamefont {M.}~\bibnamefont {Minola}}, \bibinfo {author} {\bibfnamefont
  {G.}~\bibnamefont {Christiani}}, \bibinfo {author} {\bibfnamefont
  {H.}~\bibnamefont {Habermeier}}, \bibinfo {author} {\bibfnamefont
  {G.}~\bibnamefont {Logvenov}},  \emph {et~al.},\ }\href {\doibase
  10.1038/nmat4682} {\bibfield  {journal} {\bibinfo  {journal} {Nat. Mater.}\
  }\textbf {\bibinfo {volume} {15}},\ \bibinfo {pages} {831} (\bibinfo {year}
  {2016})}\BibitemShut {NoStop}%
\bibitem [{\citenamefont {Smilde}\ \emph {et~al.}(2005)\citenamefont {Smilde},
  \citenamefont {Golubov}, \citenamefont {Ariando}, \citenamefont {Rijnders},
  \citenamefont {Dekkers}, \citenamefont {Harkema}, \citenamefont {Blank},
  \citenamefont {Rogalla},\ and\ \citenamefont
  {Hilgenkamp}}]{PhysRevLett.95.257001}%
  \BibitemOpen
  \bibfield  {author} {\bibinfo {author} {\bibfnamefont {H.~J.~H.}\
  \bibnamefont {Smilde}}, \bibinfo {author} {\bibfnamefont {A.~A.}\
  \bibnamefont {Golubov}}, \bibinfo {author} {\bibnamefont {Ariando}}, \bibinfo
  {author} {\bibfnamefont {G.}~\bibnamefont {Rijnders}}, \bibinfo {author}
  {\bibfnamefont {J.~M.}\ \bibnamefont {Dekkers}}, \bibinfo {author}
  {\bibfnamefont {S.}~\bibnamefont {Harkema}}, \bibinfo {author} {\bibfnamefont
  {D.~H.~A.}\ \bibnamefont {Blank}}, \bibinfo {author} {\bibfnamefont
  {H.}~\bibnamefont {Rogalla}}, \ and\ \bibinfo {author} {\bibfnamefont
  {H.}~\bibnamefont {Hilgenkamp}},\ }\href {\doibase
  10.1103/PhysRevLett.95.257001} {\bibfield  {journal} {\bibinfo  {journal}
  {Phys. Rev. Lett.}\ }\textbf {\bibinfo {volume} {95}},\ \bibinfo {pages}
  {257001} (\bibinfo {year} {2005})}\BibitemShut {NoStop}%
\bibitem [{\citenamefont {Dekkers}\ \emph {et~al.}(2003)\citenamefont
  {Dekkers}, \citenamefont {Rijnders}, \citenamefont {Harkema}, \citenamefont
  {Smilde}, \citenamefont {Hilgenkamp}, \citenamefont {Rogalla},\ and\
  \citenamefont {Blank}}]{Dekkers2003}%
  \BibitemOpen
  \bibfield  {author} {\bibinfo {author} {\bibfnamefont {J.~M.}\ \bibnamefont
  {Dekkers}}, \bibinfo {author} {\bibfnamefont {G.}~\bibnamefont {Rijnders}},
  \bibinfo {author} {\bibfnamefont {S.}~\bibnamefont {Harkema}}, \bibinfo
  {author} {\bibfnamefont {H.~J.~H.}\ \bibnamefont {Smilde}}, \bibinfo {author}
  {\bibfnamefont {H.}~\bibnamefont {Hilgenkamp}}, \bibinfo {author}
  {\bibfnamefont {H.}~\bibnamefont {Rogalla}}, \ and\ \bibinfo {author}
  {\bibfnamefont {D.~H.~A.}\ \bibnamefont {Blank}},\ }\href {\doibase
  10.1063/1.1633010} {\bibfield  {journal} {\bibinfo  {journal} {Appl. Phys.
  Lett.}\ }\textbf {\bibinfo {volume} {83}},\ \bibinfo {pages} {5199} (\bibinfo
  {year} {2003})}\BibitemShut {NoStop}%
\bibitem [{\citenamefont {Kwok}\ \emph {et~al.}(1991)\citenamefont {Kwok},
  \citenamefont {Welp}, \citenamefont {Vinokur}, \citenamefont {Fleshler},
  \citenamefont {Downey},\ and\ \citenamefont {Crabtree}}]{PhysRevLett.67.390}%
  \BibitemOpen
  \bibfield  {author} {\bibinfo {author} {\bibfnamefont {W.~K.}\ \bibnamefont
  {Kwok}}, \bibinfo {author} {\bibfnamefont {U.}~\bibnamefont {Welp}}, \bibinfo
  {author} {\bibfnamefont {V.~M.}\ \bibnamefont {Vinokur}}, \bibinfo {author}
  {\bibfnamefont {S.}~\bibnamefont {Fleshler}}, \bibinfo {author}
  {\bibfnamefont {J.}~\bibnamefont {Downey}}, \ and\ \bibinfo {author}
  {\bibfnamefont {G.~W.}\ \bibnamefont {Crabtree}},\ }\href {\doibase
  10.1103/PhysRevLett.67.390} {\bibfield  {journal} {\bibinfo  {journal} {Phys.
  Rev. Lett.}\ }\textbf {\bibinfo {volume} {67}},\ \bibinfo {pages} {390}
  (\bibinfo {year} {1991})}\BibitemShut {NoStop}%
\bibitem [{Note1()}]{Note1}%
  \BibitemOpen
  \bibinfo {note} {Partially detwinned films have also been achieved on (001)
  oriented NdGaO$_3$ substrates. However, the strain induced in YBCO is
  tensile, as on vicinal angle (001) oriented STO substrates \cite
  {villard1996ab}. Moreover, 45$^\circ $ grain boundaries and defects,
  influencing the transport properties, have been also reported \cite
  {scherer1992anisotropy, klemenz1999defects}. On (110) oriented NdGaO$_3$
  substrates, YBCO films grow instead twinned \cite {steinborn1994twinning,
  schweitzer1996twinning}. An exception is given by a recent work, where the
  achievement of untwinned films has been presented \cite
  {khanof2018coherently}. Here, the strain is tensile along $a$ and compressive
  along $b$, resulting in a strong suppression of orthorhombicity in the YBCO
  structure.}\BibitemShut {Stop}%
\bibitem [{\citenamefont {Br{\"o}tz}\ and\ \citenamefont
  {Fuess}(2000)}]{brotz2000detwinning}%
  \BibitemOpen
  \bibfield  {author} {\bibinfo {author} {\bibfnamefont {J.}~\bibnamefont
  {Br{\"o}tz}}\ and\ \bibinfo {author} {\bibfnamefont {H.}~\bibnamefont
  {Fuess}},\ }\href {\doibase 10.1016/S0921-4534(00)00351-8} {\bibfield
  {journal} {\bibinfo  {journal} {Physica C}\ }\textbf {\bibinfo {volume}
  {339}},\ \bibinfo {pages} {75} (\bibinfo {year} {2000})}\BibitemShut
  {NoStop}%
\bibitem [{\citenamefont {Zegenhagen}\ \emph {et~al.}(1998)\citenamefont
  {Zegenhagen}, \citenamefont {Haage},\ and\ \citenamefont
  {Jiang}}]{zegenhagen1998microscopic}%
  \BibitemOpen
  \bibfield  {author} {\bibinfo {author} {\bibfnamefont {J.}~\bibnamefont
  {Zegenhagen}}, \bibinfo {author} {\bibfnamefont {T.}~\bibnamefont {Haage}}, \
  and\ \bibinfo {author} {\bibfnamefont {Q.}~\bibnamefont {Jiang}},\ }\href
  {\doibase 10.1007/s003390050843} {\bibfield  {journal} {\bibinfo  {journal}
  {Appl. Phys. A}\ }\textbf {\bibinfo {volume} {67}},\ \bibinfo {pages} {711}
  (\bibinfo {year} {1998})}\BibitemShut {NoStop}%
\bibitem [{\citenamefont {Zhuravel}\ \emph {et~al.}(2010)\citenamefont
  {Zhuravel}, \citenamefont {Anlage}, \citenamefont {Remillard}, \citenamefont
  {Lukashenko},\ and\ \citenamefont {Ustinov}}]{zhuravel2010effect}%
  \BibitemOpen
  \bibfield  {author} {\bibinfo {author} {\bibfnamefont {A.}~\bibnamefont
  {Zhuravel}}, \bibinfo {author} {\bibfnamefont {S.~M.}\ \bibnamefont
  {Anlage}}, \bibinfo {author} {\bibfnamefont {S.~K.}\ \bibnamefont
  {Remillard}}, \bibinfo {author} {\bibfnamefont {A.}~\bibnamefont
  {Lukashenko}}, \ and\ \bibinfo {author} {\bibfnamefont {A.}~\bibnamefont
  {Ustinov}},\ }\href {\doibase 10.1063/1.3467003} {\bibfield  {journal}
  {\bibinfo  {journal} {J. Appl. Phys.}\ }\textbf {\bibinfo {volume} {108}},\
  \bibinfo {pages} {033920} (\bibinfo {year} {2010})}\BibitemShut {NoStop}%
\bibitem [{\citenamefont {Mazierska}\ \emph {et~al.}(2005)\citenamefont
  {Mazierska}, \citenamefont {Ledenyov}, \citenamefont {Jacob},\ and\
  \citenamefont {Krupka}}]{mazierska2005precise}%
  \BibitemOpen
  \bibfield  {author} {\bibinfo {author} {\bibfnamefont {J.}~\bibnamefont
  {Mazierska}}, \bibinfo {author} {\bibfnamefont {D.}~\bibnamefont {Ledenyov}},
  \bibinfo {author} {\bibfnamefont {M.~V.}\ \bibnamefont {Jacob}}, \ and\
  \bibinfo {author} {\bibfnamefont {J.}~\bibnamefont {Krupka}},\ }\href
  {\doibase 10.1088/0953-2048/18/1/004} {\bibfield  {journal} {\bibinfo
  {journal} {Supercond. Sci. Technol.}\ }\textbf {\bibinfo {volume} {18}},\
  \bibinfo {pages} {18} (\bibinfo {year} {2005})}\BibitemShut {NoStop}%
\bibitem [{\citenamefont {Scheel}\ \emph {et~al.}(1991)\citenamefont {Scheel},
  \citenamefont {Berkowski},\ and\ \citenamefont
  {Chabot}}]{scheel1991substrates}%
  \BibitemOpen
  \bibfield  {author} {\bibinfo {author} {\bibfnamefont {H.~J.}\ \bibnamefont
  {Scheel}}, \bibinfo {author} {\bibfnamefont {M.}~\bibnamefont {Berkowski}}, \
  and\ \bibinfo {author} {\bibfnamefont {B.}~\bibnamefont {Chabot}},\ }\href
  {\doibase 10.1016/0921-4534(91)91172-Z} {\bibfield  {journal} {\bibinfo
  {journal} {Physica C}\ }\textbf {\bibinfo {volume} {185}},\ \bibinfo {pages}
  {2095} (\bibinfo {year} {1991})}\BibitemShut {NoStop}%
\bibitem [{\citenamefont {Villard}\ \emph {et~al.}(1996)\citenamefont
  {Villard}, \citenamefont {Koren}, \citenamefont {Cohen}, \citenamefont
  {Polturak}, \citenamefont {Thrane},\ and\ \citenamefont
  {Chateignier}}]{villard1996ab}%
  \BibitemOpen
  \bibfield  {author} {\bibinfo {author} {\bibfnamefont {C.}~\bibnamefont
  {Villard}}, \bibinfo {author} {\bibfnamefont {G.}~\bibnamefont {Koren}},
  \bibinfo {author} {\bibfnamefont {D.}~\bibnamefont {Cohen}}, \bibinfo
  {author} {\bibfnamefont {E.}~\bibnamefont {Polturak}}, \bibinfo {author}
  {\bibfnamefont {B.}~\bibnamefont {Thrane}}, \ and\ \bibinfo {author}
  {\bibfnamefont {D.}~\bibnamefont {Chateignier}},\ }\href {\doibase
  10.1103/PhysRevLett.77.3913} {\bibfield  {journal} {\bibinfo  {journal}
  {Phys. Rev. Lett.}\ }\textbf {\bibinfo {volume} {77}},\ \bibinfo {pages}
  {3913} (\bibinfo {year} {1996})}\BibitemShut {NoStop}%
\bibitem [{\citenamefont {Dubourdieu}\ \emph {et~al.}(1996)\citenamefont
  {Dubourdieu}, \citenamefont {Senateur}, \citenamefont {Thomas}, \citenamefont
  {Weiss}, \citenamefont {Thrane},\ and\ \citenamefont
  {Brunel}}]{dubourdieu1996twinning}%
  \BibitemOpen
  \bibfield  {author} {\bibinfo {author} {\bibfnamefont {C.}~\bibnamefont
  {Dubourdieu}}, \bibinfo {author} {\bibfnamefont {J.}~\bibnamefont
  {Senateur}}, \bibinfo {author} {\bibfnamefont {O.}~\bibnamefont {Thomas}},
  \bibinfo {author} {\bibfnamefont {F.}~\bibnamefont {Weiss}}, \bibinfo
  {author} {\bibfnamefont {B.}~\bibnamefont {Thrane}}, \ and\ \bibinfo {author}
  {\bibfnamefont {M.}~\bibnamefont {Brunel}},\ }\href {\doibase
  10.1063/1.117629} {\bibfield  {journal} {\bibinfo  {journal} {Appl. Phys.
  Lett.}\ }\textbf {\bibinfo {volume} {69}},\ \bibinfo {pages} {1942} (\bibinfo
  {year} {1996})}\BibitemShut {NoStop}%
\bibitem [{\citenamefont {Bdikin}\ \emph {et~al.}(2002)\citenamefont {Bdikin},
  \citenamefont {Mozhaev}, \citenamefont {Ovsyannikov}, \citenamefont
  {Komissinski},\ and\ \citenamefont {Kotelyanskii}}]{bdikin2002growth}%
  \BibitemOpen
  \bibfield  {author} {\bibinfo {author} {\bibfnamefont {I.}~\bibnamefont
  {Bdikin}}, \bibinfo {author} {\bibfnamefont {P.}~\bibnamefont {Mozhaev}},
  \bibinfo {author} {\bibfnamefont {G.}~\bibnamefont {Ovsyannikov}}, \bibinfo
  {author} {\bibfnamefont {P.}~\bibnamefont {Komissinski}}, \ and\ \bibinfo
  {author} {\bibfnamefont {I.}~\bibnamefont {Kotelyanskii}},\ }\href {\doibase
  10.1016/S0921-4534(01)01106-6} {\bibfield  {journal} {\bibinfo  {journal}
  {Physica C}\ }\textbf {\bibinfo {volume} {377}},\ \bibinfo {pages} {26}
  (\bibinfo {year} {2002})}\BibitemShut {NoStop}%
\bibitem [{\citenamefont {Nam}\ \emph {et~al.}(2005)\citenamefont {Nam},
  \citenamefont {Hughes}, \citenamefont {Castellan}, \citenamefont {Gaulin},
  \citenamefont {Britten},\ and\ \citenamefont {Preston}}]{nam2005origin}%
  \BibitemOpen
  \bibfield  {author} {\bibinfo {author} {\bibfnamefont {J.}~\bibnamefont
  {Nam}}, \bibinfo {author} {\bibfnamefont {R.}~\bibnamefont {Hughes}},
  \bibinfo {author} {\bibfnamefont {J.}~\bibnamefont {Castellan}}, \bibinfo
  {author} {\bibfnamefont {B.}~\bibnamefont {Gaulin}}, \bibinfo {author}
  {\bibfnamefont {J.}~\bibnamefont {Britten}}, \ and\ \bibinfo {author}
  {\bibfnamefont {J.}~\bibnamefont {Preston}},\ }\href {\doibase
  10.1063/1.1926395} {\bibfield  {journal} {\bibinfo  {journal} {J. Appl.
  Phys.}\ }\textbf {\bibinfo {volume} {97}},\ \bibinfo {pages} {123906}
  (\bibinfo {year} {2005})}\BibitemShut {NoStop}%
\bibitem [{\citenamefont {Salluzzo}\ \emph {et~al.}(2005)\citenamefont
  {Salluzzo}, \citenamefont {De~Luca}, \citenamefont {Marr{\`e}}, \citenamefont
  {Putti}, \citenamefont {Tropeano}, \citenamefont {Di~Uccio},\ and\
  \citenamefont {Vaglio}}]{salluzzo2005thickness}%
  \BibitemOpen
  \bibfield  {author} {\bibinfo {author} {\bibfnamefont {M.}~\bibnamefont
  {Salluzzo}}, \bibinfo {author} {\bibfnamefont {G.}~\bibnamefont {De~Luca}},
  \bibinfo {author} {\bibfnamefont {D.}~\bibnamefont {Marr{\`e}}}, \bibinfo
  {author} {\bibfnamefont {M.}~\bibnamefont {Putti}}, \bibinfo {author}
  {\bibfnamefont {M.}~\bibnamefont {Tropeano}}, \bibinfo {author}
  {\bibfnamefont {U.~S.}\ \bibnamefont {Di~Uccio}}, \ and\ \bibinfo {author}
  {\bibfnamefont {R.}~\bibnamefont {Vaglio}},\ }\href {\doibase
  10.1103/PhysRevB.72.134521} {\bibfield  {journal} {\bibinfo  {journal} {Phys.
  Rev. B}\ }\textbf {\bibinfo {volume} {72}},\ \bibinfo {pages} {134521}
  (\bibinfo {year} {2005})}\BibitemShut {NoStop}%
\bibitem [{\citenamefont {Ho}\ and\ \citenamefont
  {Bohnen}(1987)}]{ho1987stability}%
  \BibitemOpen
  \bibfield  {author} {\bibinfo {author} {\bibfnamefont {K.-M.}\ \bibnamefont
  {Ho}}\ and\ \bibinfo {author} {\bibfnamefont {K.}~\bibnamefont {Bohnen}},\
  }\href {\doibase 10.1103/PhysRevLett.59.1833} {\bibfield  {journal} {\bibinfo
   {journal} {Phys. Rev. Lett.}\ }\textbf {\bibinfo {volume} {59}},\ \bibinfo
  {pages} {1833} (\bibinfo {year} {1987})}\BibitemShut {NoStop}%
\bibitem [{\citenamefont {Gritsch}\ \emph {et~al.}(1989)\citenamefont
  {Gritsch}, \citenamefont {Coulman}, \citenamefont {Behm},\ and\ \citenamefont
  {Ertl}}]{gritsch1989mechanism}%
  \BibitemOpen
  \bibfield  {author} {\bibinfo {author} {\bibfnamefont {T.}~\bibnamefont
  {Gritsch}}, \bibinfo {author} {\bibfnamefont {D.}~\bibnamefont {Coulman}},
  \bibinfo {author} {\bibfnamefont {R.}~\bibnamefont {Behm}}, \ and\ \bibinfo
  {author} {\bibfnamefont {G.}~\bibnamefont {Ertl}},\ }\href {\doibase
  10.1103/PhysRevLett.63.1086} {\bibfield  {journal} {\bibinfo  {journal}
  {Phys. Rev. Lett.}\ }\textbf {\bibinfo {volume} {63}},\ \bibinfo {pages}
  {1086} (\bibinfo {year} {1989})}\BibitemShut {NoStop}%
\bibitem [{\citenamefont {Koch}\ \emph {et~al.}(1991)\citenamefont {Koch},
  \citenamefont {Borbonus}, \citenamefont {Haase},\ and\ \citenamefont
  {Rieder}}]{koch1991new}%
  \BibitemOpen
  \bibfield  {author} {\bibinfo {author} {\bibfnamefont {R.}~\bibnamefont
  {Koch}}, \bibinfo {author} {\bibfnamefont {M.}~\bibnamefont {Borbonus}},
  \bibinfo {author} {\bibfnamefont {O.}~\bibnamefont {Haase}}, \ and\ \bibinfo
  {author} {\bibfnamefont {K.}~\bibnamefont {Rieder}},\ }\href {\doibase
  10.1103/PhysRevLett.67.3416} {\bibfield  {journal} {\bibinfo  {journal}
  {Phys. Rev. Lett.}\ }\textbf {\bibinfo {volume} {67}},\ \bibinfo {pages}
  {3416} (\bibinfo {year} {1991})}\BibitemShut {NoStop}%
\bibitem [{\citenamefont {Henrich}(1976)}]{henrich1976thermal}%
  \BibitemOpen
  \bibfield  {author} {\bibinfo {author} {\bibfnamefont {V.~E.}\ \bibnamefont
  {Henrich}},\ }\href {\doibase 10.1016/0039-6028(76)90190-4} {\bibfield
  {journal} {\bibinfo  {journal} {Surf. Sci.}\ }\textbf {\bibinfo {volume}
  {57}},\ \bibinfo {pages} {385} (\bibinfo {year} {1976})}\BibitemShut
  {NoStop}%
\bibitem [{\citenamefont {De~Leeuw}\ \emph {et~al.}(1995)\citenamefont
  {De~Leeuw}, \citenamefont {Watson},\ and\ \citenamefont
  {Parker}}]{de1995atomistic}%
  \BibitemOpen
  \bibfield  {author} {\bibinfo {author} {\bibfnamefont {N.}~\bibnamefont
  {De~Leeuw}}, \bibinfo {author} {\bibfnamefont {G.}~\bibnamefont {Watson}}, \
  and\ \bibinfo {author} {\bibfnamefont {S.}~\bibnamefont {Parker}},\ }\href
  {\doibase 10.1021/j100047a028} {\bibfield  {journal} {\bibinfo  {journal} {J.
  Phys. Chem.}\ }\textbf {\bibinfo {volume} {99}},\ \bibinfo {pages} {17219}
  (\bibinfo {year} {1995})}\BibitemShut {NoStop}%
\bibitem [{\citenamefont {Giese}\ \emph {et~al.}(2000)\citenamefont {Giese},
  \citenamefont {Lamelas}, \citenamefont {Owen}, \citenamefont {Plass},\ and\
  \citenamefont {Gajdardziska-Josifovska}}]{giese2000atomic}%
  \BibitemOpen
  \bibfield  {author} {\bibinfo {author} {\bibfnamefont {D.}~\bibnamefont
  {Giese}}, \bibinfo {author} {\bibfnamefont {F.}~\bibnamefont {Lamelas}},
  \bibinfo {author} {\bibfnamefont {H.}~\bibnamefont {Owen}}, \bibinfo {author}
  {\bibfnamefont {R.}~\bibnamefont {Plass}}, \ and\ \bibinfo {author}
  {\bibfnamefont {M.}~\bibnamefont {Gajdardziska-Josifovska}},\ }\href
  {\doibase 10.1016/S0039-6028(00)00382-4} {\bibfield  {journal} {\bibinfo
  {journal} {Surf. Sci.}\ }\textbf {\bibinfo {volume} {457}},\ \bibinfo {pages}
  {326} (\bibinfo {year} {2000})}\BibitemShut {NoStop}%
\bibitem [{\citenamefont {Chern}\ \emph {et~al.}(1998)\citenamefont {Chern},
  \citenamefont {Huang},\ and\ \citenamefont {Leung}}]{chern1998atomic}%
  \BibitemOpen
  \bibfield  {author} {\bibinfo {author} {\bibfnamefont {G.}~\bibnamefont
  {Chern}}, \bibinfo {author} {\bibfnamefont {J.}~\bibnamefont {Huang}}, \ and\
  \bibinfo {author} {\bibfnamefont {T.}~\bibnamefont {Leung}},\ }\href
  {\doibase 10.1116/1.581279} {\bibfield  {journal} {\bibinfo  {journal} {J.
  Vac. Sci. Technol. A}\ }\textbf {\bibinfo {volume} {16}},\ \bibinfo {pages}
  {964} (\bibinfo {year} {1998})}\BibitemShut {NoStop}%
\bibitem [{\citenamefont {Sugawara}\ and\ \citenamefont
  {Mae}(2004)}]{sugawara2004faceting}%
  \BibitemOpen
  \bibfield  {author} {\bibinfo {author} {\bibfnamefont {A.}~\bibnamefont
  {Sugawara}}\ and\ \bibinfo {author} {\bibfnamefont {K.}~\bibnamefont {Mae}},\
  }\href {\doibase 10.1016/j.susc.2004.04.002} {\bibfield  {journal} {\bibinfo
  {journal} {Surf. Sci.}\ }\textbf {\bibinfo {volume} {558}},\ \bibinfo {pages}
  {211} (\bibinfo {year} {2004})}\BibitemShut {NoStop}%
\bibitem [{\citenamefont {Ciancio}\ \emph {et~al.}(2013)\citenamefont
  {Ciancio}, \citenamefont {Vittadini}, \citenamefont {Selloni}, \citenamefont
  {Arpaia}, \citenamefont {Aruta}, \citenamefont {Granozio}, \citenamefont
  {Di~Uccio}, \citenamefont {Rossi},\ and\ \citenamefont
  {Carlino}}]{ciancio2013evolution}%
  \BibitemOpen
  \bibfield  {author} {\bibinfo {author} {\bibfnamefont {R.}~\bibnamefont
  {Ciancio}}, \bibinfo {author} {\bibfnamefont {A.}~\bibnamefont {Vittadini}},
  \bibinfo {author} {\bibfnamefont {A.}~\bibnamefont {Selloni}}, \bibinfo
  {author} {\bibfnamefont {R.}~\bibnamefont {Arpaia}}, \bibinfo {author}
  {\bibfnamefont {C.}~\bibnamefont {Aruta}}, \bibinfo {author} {\bibfnamefont
  {F.~M.}\ \bibnamefont {Granozio}}, \bibinfo {author} {\bibfnamefont {U.~S.}\
  \bibnamefont {Di~Uccio}}, \bibinfo {author} {\bibfnamefont {G.}~\bibnamefont
  {Rossi}}, \ and\ \bibinfo {author} {\bibfnamefont {E.}~\bibnamefont
  {Carlino}},\ }\href {\doibase 10.1007/s11051-013-1735-x} {\bibfield
  {journal} {\bibinfo  {journal} {J. Nanoparticle Res.}\ }\textbf {\bibinfo
  {volume} {15}},\ \bibinfo {pages} {1735} (\bibinfo {year}
  {2013})}\BibitemShut {NoStop}%
\bibitem [{\citenamefont {Dion}\ \emph {et~al.}(2004)\citenamefont {Dion},
  \citenamefont {Rydberg}, \citenamefont {Schr{\"o}der}, \citenamefont
  {Langreth},\ and\ \citenamefont {Lundqvist}}]{dion2004van}%
  \BibitemOpen
  \bibfield  {author} {\bibinfo {author} {\bibfnamefont {M.}~\bibnamefont
  {Dion}}, \bibinfo {author} {\bibfnamefont {H.}~\bibnamefont {Rydberg}},
  \bibinfo {author} {\bibfnamefont {E.}~\bibnamefont {Schr{\"o}der}}, \bibinfo
  {author} {\bibfnamefont {D.~C.}\ \bibnamefont {Langreth}}, \ and\ \bibinfo
  {author} {\bibfnamefont {B.~I.}\ \bibnamefont {Lundqvist}},\ }\href {\doibase
  10.1103/PhysRevLett.92.246401} {\bibfield  {journal} {\bibinfo  {journal}
  {Phys. Rev. Lett.}\ }\textbf {\bibinfo {volume} {92}},\ \bibinfo {pages}
  {246401} (\bibinfo {year} {2004})}\BibitemShut {NoStop}%
\bibitem [{\citenamefont {Thonhauser}\ \emph {et~al.}(2007)\citenamefont
  {Thonhauser}, \citenamefont {Cooper}, \citenamefont {Li}, \citenamefont
  {Puzder}, \citenamefont {Hyldgaard},\ and\ \citenamefont
  {Langreth}}]{thonhauser2007van}%
  \BibitemOpen
  \bibfield  {author} {\bibinfo {author} {\bibfnamefont {T.}~\bibnamefont
  {Thonhauser}}, \bibinfo {author} {\bibfnamefont {V.~R.}\ \bibnamefont
  {Cooper}}, \bibinfo {author} {\bibfnamefont {S.}~\bibnamefont {Li}}, \bibinfo
  {author} {\bibfnamefont {A.}~\bibnamefont {Puzder}}, \bibinfo {author}
  {\bibfnamefont {P.}~\bibnamefont {Hyldgaard}}, \ and\ \bibinfo {author}
  {\bibfnamefont {D.~C.}\ \bibnamefont {Langreth}},\ }\href {\doibase
  10.1103/PhysRevB.76.125112} {\bibfield  {journal} {\bibinfo  {journal} {Phys.
  Rev. B}\ }\textbf {\bibinfo {volume} {76}},\ \bibinfo {pages} {125112}
  (\bibinfo {year} {2007})}\BibitemShut {NoStop}%
\bibitem [{\citenamefont {Lee}\ \emph {et~al.}(2010)\citenamefont {Lee},
  \citenamefont {Murray}, \citenamefont {Kong}, \citenamefont {Lundqvist},\
  and\ \citenamefont {Langreth}}]{lee2010higher}%
  \BibitemOpen
  \bibfield  {author} {\bibinfo {author} {\bibfnamefont {K.}~\bibnamefont
  {Lee}}, \bibinfo {author} {\bibfnamefont {{\'E}.~D.}\ \bibnamefont {Murray}},
  \bibinfo {author} {\bibfnamefont {L.}~\bibnamefont {Kong}}, \bibinfo {author}
  {\bibfnamefont {B.~I.}\ \bibnamefont {Lundqvist}}, \ and\ \bibinfo {author}
  {\bibfnamefont {D.~C.}\ \bibnamefont {Langreth}},\ }\href {\doibase
  10.1103/PhysRevB.82.081101} {\bibfield  {journal} {\bibinfo  {journal} {Phys.
  Rev. B}\ }\textbf {\bibinfo {volume} {82}},\ \bibinfo {pages} {081101(R)}
  (\bibinfo {year} {2010})}\BibitemShut {NoStop}%
\bibitem [{\citenamefont {Berland}\ \emph {et~al.}(2015)\citenamefont
  {Berland}, \citenamefont {Cooper}, \citenamefont {Lee}, \citenamefont
  {Schr{\"o}der}, \citenamefont {Thonhauser}, \citenamefont {Hyldgaard},\ and\
  \citenamefont {Lundqvist}}]{berland2015van}%
  \BibitemOpen
  \bibfield  {author} {\bibinfo {author} {\bibfnamefont {K.}~\bibnamefont
  {Berland}}, \bibinfo {author} {\bibfnamefont {V.~R.}\ \bibnamefont {Cooper}},
  \bibinfo {author} {\bibfnamefont {K.}~\bibnamefont {Lee}}, \bibinfo {author}
  {\bibfnamefont {E.}~\bibnamefont {Schr{\"o}der}}, \bibinfo {author}
  {\bibfnamefont {T.}~\bibnamefont {Thonhauser}}, \bibinfo {author}
  {\bibfnamefont {P.}~\bibnamefont {Hyldgaard}}, \ and\ \bibinfo {author}
  {\bibfnamefont {B.~I.}\ \bibnamefont {Lundqvist}},\ }\href {\doibase
  10.1088/0034-4885/78/6/066501} {\bibfield  {journal} {\bibinfo  {journal}
  {Rep. Prog. Phys.}\ }\textbf {\bibinfo {volume} {78}},\ \bibinfo {pages}
  {066501} (\bibinfo {year} {2015})}\BibitemShut {NoStop}%
\bibitem [{\citenamefont {Berland}\ and\ \citenamefont
  {Hyldgaard}(2014)}]{berland2014exchange}%
  \BibitemOpen
  \bibfield  {author} {\bibinfo {author} {\bibfnamefont {K.}~\bibnamefont
  {Berland}}\ and\ \bibinfo {author} {\bibfnamefont {P.}~\bibnamefont
  {Hyldgaard}},\ }\href {\doibase 10.1103/PhysRevB.89.035412} {\bibfield
  {journal} {\bibinfo  {journal} {Phys. Rev. B}\ }\textbf {\bibinfo {volume}
  {89}},\ \bibinfo {pages} {035412} (\bibinfo {year} {2014})}\BibitemShut
  {NoStop}%
\bibitem [{\citenamefont {Giannozzi}\ \emph {et~al.}(2009)\citenamefont
  {Giannozzi}, \citenamefont {Baroni}, \citenamefont {Bonini}, \citenamefont
  {Calandra}, \citenamefont {Car}, \citenamefont {Cavazzoni}, \citenamefont
  {Ceresoli}, \citenamefont {Chiarotti}, \citenamefont {Cococcioni},
  \citenamefont {Dabo} \emph {et~al.}}]{giannozzi2009quantum}%
  \BibitemOpen
  \bibfield  {author} {\bibinfo {author} {\bibfnamefont {P.}~\bibnamefont
  {Giannozzi}}, \bibinfo {author} {\bibfnamefont {S.}~\bibnamefont {Baroni}},
  \bibinfo {author} {\bibfnamefont {N.}~\bibnamefont {Bonini}}, \bibinfo
  {author} {\bibfnamefont {M.}~\bibnamefont {Calandra}}, \bibinfo {author}
  {\bibfnamefont {R.}~\bibnamefont {Car}}, \bibinfo {author} {\bibfnamefont
  {C.}~\bibnamefont {Cavazzoni}}, \bibinfo {author} {\bibfnamefont
  {D.}~\bibnamefont {Ceresoli}}, \bibinfo {author} {\bibfnamefont {G.~L.}\
  \bibnamefont {Chiarotti}}, \bibinfo {author} {\bibfnamefont {M.}~\bibnamefont
  {Cococcioni}}, \bibinfo {author} {\bibfnamefont {I.}~\bibnamefont {Dabo}},
  \emph {et~al.},\ }\href {\doibase 10.1088/0953-8984/21/39/395502} {\bibfield
  {journal} {\bibinfo  {journal} {J. Phys.: Condens. Matter}\ }\textbf
  {\bibinfo {volume} {21}},\ \bibinfo {pages} {395502} (\bibinfo {year}
  {2009})}\BibitemShut {NoStop}%
\bibitem [{\citenamefont {Rom{\'a}n-P{\'e}rez}\ and\ \citenamefont
  {Soler}(2009)}]{roman2009efficient}%
  \BibitemOpen
  \bibfield  {author} {\bibinfo {author} {\bibfnamefont {G.}~\bibnamefont
  {Rom{\'a}n-P{\'e}rez}}\ and\ \bibinfo {author} {\bibfnamefont {J.~M.}\
  \bibnamefont {Soler}},\ }\href {\doibase 10.1103/PhysRevLett.103.096102}
  {\bibfield  {journal} {\bibinfo  {journal} {Phys. Rev. Lett.}\ }\textbf
  {\bibinfo {volume} {103}},\ \bibinfo {pages} {096102} (\bibinfo {year}
  {2009})}\BibitemShut {NoStop}%
\bibitem [{\citenamefont {Arpaia}(2016)}]{arpaia2016yba2cu3o7}%
  \BibitemOpen
  \bibfield  {author} {\bibinfo {author} {\bibfnamefont {R.}~\bibnamefont
  {Arpaia}},\ }\href@noop {} {\emph {\bibinfo {title}
  {YBa$_2$Cu$_3$O$_{7-\delta}$ Nanowires to Study Nanoscale Ordering in
  High-T$_C$ Superconductors}}}\ (\bibinfo  {publisher} {PhD thesis, Department
  of Microtechnology and Nanoscience, Chalmers University of Technology},\
  \bibinfo {year} {2016})\BibitemShut {NoStop}%
\bibitem [{Note2()}]{Note2}%
  \BibitemOpen
  \bibinfo {note} {Slightly overdoped films are the best systems where to study
  the twinning state, since at this oxygen doping level the in-plane
  orthorhombicity of the YBCO unit cell, i.e. the difference in length between
  the $a$ and the $b$ axes, is strongest \cite
  {ye1993quantitative}.}\BibitemShut {Stop}%
\bibitem [{\citenamefont {Baghdadi}\ \emph
  {et~al.}(2015{\natexlab{a}})\citenamefont {Baghdadi}, \citenamefont {Arpaia},
  \citenamefont {Bauch},\ and\ \citenamefont {Lombardi}}]{baghdadi2015toward}%
  \BibitemOpen
  \bibfield  {author} {\bibinfo {author} {\bibfnamefont {R.}~\bibnamefont
  {Baghdadi}}, \bibinfo {author} {\bibfnamefont {R.}~\bibnamefont {Arpaia}},
  \bibinfo {author} {\bibfnamefont {T.}~\bibnamefont {Bauch}}, \ and\ \bibinfo
  {author} {\bibfnamefont {F.}~\bibnamefont {Lombardi}},\ }\href {\doibase
  10.1109/TASC.2014.2362991} {\bibfield  {journal} {\bibinfo  {journal} {IEEE
  Trans. Appl. Supercond.}\ }\textbf {\bibinfo {volume} {25}},\ \bibinfo
  {pages} {1100104} (\bibinfo {year} {2015}{\natexlab{a}})}\BibitemShut
  {NoStop}%
\bibitem [{\citenamefont {Arpaia}\ \emph {et~al.}(2017)\citenamefont {Arpaia},
  \citenamefont {Golubev}, \citenamefont {Baghdadi}, \citenamefont {Ciancio},
  \citenamefont {Dra{\v{z}}i{\'c}}, \citenamefont {Orgiani}, \citenamefont
  {Montemurro}, \citenamefont {Bauch},\ and\ \citenamefont
  {Lombardi}}]{arpaia2017transport}%
  \BibitemOpen
  \bibfield  {author} {\bibinfo {author} {\bibfnamefont {R.}~\bibnamefont
  {Arpaia}}, \bibinfo {author} {\bibfnamefont {D.}~\bibnamefont {Golubev}},
  \bibinfo {author} {\bibfnamefont {R.}~\bibnamefont {Baghdadi}}, \bibinfo
  {author} {\bibfnamefont {R.}~\bibnamefont {Ciancio}}, \bibinfo {author}
  {\bibfnamefont {G.}~\bibnamefont {Dra{\v{z}}i{\'c}}}, \bibinfo {author}
  {\bibfnamefont {P.}~\bibnamefont {Orgiani}}, \bibinfo {author} {\bibfnamefont
  {D.}~\bibnamefont {Montemurro}}, \bibinfo {author} {\bibfnamefont
  {T.}~\bibnamefont {Bauch}}, \ and\ \bibinfo {author} {\bibfnamefont
  {F.}~\bibnamefont {Lombardi}},\ }\href {\doibase 10.1103/PhysRevB.96.064525}
  {\bibfield  {journal} {\bibinfo  {journal} {Phys. Rev. B}\ }\textbf {\bibinfo
  {volume} {96}},\ \bibinfo {pages} {064525} (\bibinfo {year}
  {2017})}\BibitemShut {NoStop}%
\bibitem [{\citenamefont {Arpaia}\ \emph {et~al.}(2018)\citenamefont {Arpaia},
  \citenamefont {Andersson}, \citenamefont {Trabaldo}, \citenamefont {Bauch},\
  and\ \citenamefont {Lombardi}}]{arpaia2018probing}%
  \BibitemOpen
  \bibfield  {author} {\bibinfo {author} {\bibfnamefont {R.}~\bibnamefont
  {Arpaia}}, \bibinfo {author} {\bibfnamefont {E.}~\bibnamefont {Andersson}},
  \bibinfo {author} {\bibfnamefont {E.}~\bibnamefont {Trabaldo}}, \bibinfo
  {author} {\bibfnamefont {T.}~\bibnamefont {Bauch}}, \ and\ \bibinfo {author}
  {\bibfnamefont {F.}~\bibnamefont {Lombardi}},\ }\href {\doibase
  10.1103/PhysRevMaterials.2.024804} {\bibfield  {journal} {\bibinfo  {journal}
  {Phys. Rev. Materials}\ }\textbf {\bibinfo {volume} {2}},\ \bibinfo {pages}
  {024804} (\bibinfo {year} {2018})}\BibitemShut {NoStop}%
\bibitem [{\citenamefont
  {Woodward}(1997{\natexlab{a}})}]{woodward1997octahedral1}%
  \BibitemOpen
  \bibfield  {author} {\bibinfo {author} {\bibfnamefont {P.~M.}\ \bibnamefont
  {Woodward}},\ }\href {\doibase 10.1107/S0108768196010713} {\bibfield
  {journal} {\bibinfo  {journal} {Acta Crystallogr. Sect. B}\ }\textbf
  {\bibinfo {volume} {53}},\ \bibinfo {pages} {32} (\bibinfo {year}
  {1997}{\natexlab{a}})}\BibitemShut {NoStop}%
\bibitem [{\citenamefont
  {Woodward}(1997{\natexlab{b}})}]{woodward1997octahedral2}%
  \BibitemOpen
  \bibfield  {author} {\bibinfo {author} {\bibfnamefont {P.~M.}\ \bibnamefont
  {Woodward}},\ }\href {\doibase 10.1107/S0108768196012050} {\bibfield
  {journal} {\bibinfo  {journal} {Acta Crystallogr. Sect. B}\ }\textbf
  {\bibinfo {volume} {53}},\ \bibinfo {pages} {44} (\bibinfo {year}
  {1997}{\natexlab{b}})}\BibitemShut {NoStop}%
\bibitem [{\citenamefont {Gebhardt}\ \emph {et~al.}(2007)\citenamefont
  {Gebhardt}, \citenamefont {Kasper}, \citenamefont {Vigliante}, \citenamefont
  {Wochner}, \citenamefont {Dosch}, \citenamefont {Razavi},\ and\ \citenamefont
  {Habermeier}}]{gebhardt2007formation}%
  \BibitemOpen
  \bibfield  {author} {\bibinfo {author} {\bibfnamefont {U.}~\bibnamefont
  {Gebhardt}}, \bibinfo {author} {\bibfnamefont {N.}~\bibnamefont {Kasper}},
  \bibinfo {author} {\bibfnamefont {A.}~\bibnamefont {Vigliante}}, \bibinfo
  {author} {\bibfnamefont {P.}~\bibnamefont {Wochner}}, \bibinfo {author}
  {\bibfnamefont {H.}~\bibnamefont {Dosch}}, \bibinfo {author} {\bibfnamefont
  {F.}~\bibnamefont {Razavi}}, \ and\ \bibinfo {author} {\bibfnamefont {H.-U.}\
  \bibnamefont {Habermeier}},\ }\href {\doibase 10.1103/PhysRevLett.98.096101}
  {\bibfield  {journal} {\bibinfo  {journal} {Phys. Rev. Lett.}\ }\textbf
  {\bibinfo {volume} {98}},\ \bibinfo {pages} {096101} (\bibinfo {year}
  {2007})}\BibitemShut {NoStop}%
\bibitem [{\citenamefont {Vailionis}\ \emph {et~al.}(2009)\citenamefont
  {Vailionis}, \citenamefont {Boschker}, \citenamefont {Houwman}, \citenamefont
  {Koster}, \citenamefont {Rijnders},\ and\ \citenamefont
  {Blank}}]{vailionis2009anisotropic}%
  \BibitemOpen
  \bibfield  {author} {\bibinfo {author} {\bibfnamefont {A.}~\bibnamefont
  {Vailionis}}, \bibinfo {author} {\bibfnamefont {H.}~\bibnamefont {Boschker}},
  \bibinfo {author} {\bibfnamefont {E.}~\bibnamefont {Houwman}}, \bibinfo
  {author} {\bibfnamefont {G.}~\bibnamefont {Koster}}, \bibinfo {author}
  {\bibfnamefont {G.}~\bibnamefont {Rijnders}}, \ and\ \bibinfo {author}
  {\bibfnamefont {D.~H.}\ \bibnamefont {Blank}},\ }\href {\doibase
  10.1063/1.3249583} {\bibfield  {journal} {\bibinfo  {journal} {Appl. Phys.
  Lett.}\ }\textbf {\bibinfo {volume} {95}},\ \bibinfo {pages} {152508}
  (\bibinfo {year} {2009})}\BibitemShut {NoStop}%
\bibitem [{\citenamefont {Vailionis}\ \emph {et~al.}(2011)\citenamefont
  {Vailionis}, \citenamefont {Boschker}, \citenamefont {Siemons}, \citenamefont
  {Houwman}, \citenamefont {Blank}, \citenamefont {Rijnders},\ and\
  \citenamefont {Koster}}]{vailionis2011misfit}%
  \BibitemOpen
  \bibfield  {author} {\bibinfo {author} {\bibfnamefont {A.}~\bibnamefont
  {Vailionis}}, \bibinfo {author} {\bibfnamefont {H.}~\bibnamefont {Boschker}},
  \bibinfo {author} {\bibfnamefont {W.}~\bibnamefont {Siemons}}, \bibinfo
  {author} {\bibfnamefont {E.~P.}\ \bibnamefont {Houwman}}, \bibinfo {author}
  {\bibfnamefont {D.~H.}\ \bibnamefont {Blank}}, \bibinfo {author}
  {\bibfnamefont {G.}~\bibnamefont {Rijnders}}, \ and\ \bibinfo {author}
  {\bibfnamefont {G.}~\bibnamefont {Koster}},\ }\href {\doibase
  10.1103/PhysRevB.83.064101} {\bibfield  {journal} {\bibinfo  {journal} {Phys.
  Rev. B}\ }\textbf {\bibinfo {volume} {83}},\ \bibinfo {pages} {064101}
  (\bibinfo {year} {2011})}\BibitemShut {NoStop}%
\bibitem [{\citenamefont {Weber}\ \emph {et~al.}(2016)\citenamefont {Weber},
  \citenamefont {Guennou}, \citenamefont {Dix}, \citenamefont {Pesquera},
  \citenamefont {S{\'a}nchez}, \citenamefont {Herranz}, \citenamefont
  {Fontcuberta}, \citenamefont {L{\'o}pez-Conesa}, \citenamefont {Estrad{\'e}},
  \citenamefont {Peir{\'o}}, , \citenamefont {I\~niguez},\ and\ \citenamefont
  {Kreisel}}]{weber2016multiple}%
  \BibitemOpen
  \bibfield  {author} {\bibinfo {author} {\bibfnamefont {M.}~\bibnamefont
  {Weber}}, \bibinfo {author} {\bibfnamefont {M.}~\bibnamefont {Guennou}},
  \bibinfo {author} {\bibfnamefont {N.}~\bibnamefont {Dix}}, \bibinfo {author}
  {\bibfnamefont {D.}~\bibnamefont {Pesquera}}, \bibinfo {author}
  {\bibfnamefont {F.}~\bibnamefont {S{\'a}nchez}}, \bibinfo {author}
  {\bibfnamefont {G.}~\bibnamefont {Herranz}}, \bibinfo {author} {\bibfnamefont
  {J.}~\bibnamefont {Fontcuberta}}, \bibinfo {author} {\bibfnamefont
  {L.}~\bibnamefont {L{\'o}pez-Conesa}}, \bibinfo {author} {\bibfnamefont
  {S.}~\bibnamefont {Estrad{\'e}}}, \bibinfo {author} {\bibfnamefont
  {F.}~\bibnamefont {Peir{\'o}}}, , \bibinfo {author} {\bibfnamefont
  {J.}~\bibnamefont {I\~niguez}}, \ and\ \bibinfo {author} {\bibfnamefont
  {J.}~\bibnamefont {Kreisel}},\ }\href {\doibase 10.1103/PhysRevB.94.014118}
  {\bibfield  {journal} {\bibinfo  {journal} {Phys. Rev. B}\ }\textbf {\bibinfo
  {volume} {94}},\ \bibinfo {pages} {014118} (\bibinfo {year}
  {2016})}\BibitemShut {NoStop}%
\bibitem [{\citenamefont {Chisholm}\ and\ \citenamefont
  {Pennycook}(1991)}]{chisholm1991structural}%
  \BibitemOpen
  \bibfield  {author} {\bibinfo {author} {\bibfnamefont {M.}~\bibnamefont
  {Chisholm}}\ and\ \bibinfo {author} {\bibfnamefont {S.}~\bibnamefont
  {Pennycook}},\ }\href {\doibase 10.1038/351047a0} {\bibfield  {journal}
  {\bibinfo  {journal} {Nature}\ }\textbf {\bibinfo {volume} {351}},\ \bibinfo
  {pages} {47} (\bibinfo {year} {1991})}\BibitemShut {NoStop}%
\bibitem [{\citenamefont {Lombardi}\ \emph {et~al.}(1998)\citenamefont
  {Lombardi}, \citenamefont {Ivanov}, \citenamefont {Fischer}, \citenamefont
  {Olsson},\ and\ \citenamefont {Claeson}}]{lombardi1998transport}%
  \BibitemOpen
  \bibfield  {author} {\bibinfo {author} {\bibfnamefont {F.}~\bibnamefont
  {Lombardi}}, \bibinfo {author} {\bibfnamefont {Z.}~\bibnamefont {Ivanov}},
  \bibinfo {author} {\bibfnamefont {G.}~\bibnamefont {Fischer}}, \bibinfo
  {author} {\bibfnamefont {E.}~\bibnamefont {Olsson}}, \ and\ \bibinfo {author}
  {\bibfnamefont {T.}~\bibnamefont {Claeson}},\ }\href {\doibase
  10.1063/1.120700} {\bibfield  {journal} {\bibinfo  {journal} {Appl. Phys.
  Lett.}\ }\textbf {\bibinfo {volume} {72}},\ \bibinfo {pages} {249} (\bibinfo
  {year} {1998})}\BibitemShut {NoStop}%
\bibitem [{\citenamefont {Arpaia}\ \emph {et~al.}(2013)\citenamefont {Arpaia},
  \citenamefont {Nawaz}, \citenamefont {Lombardi},\ and\ \citenamefont
  {Bauch}}]{arpaia2013improved}%
  \BibitemOpen
  \bibfield  {author} {\bibinfo {author} {\bibfnamefont {R.}~\bibnamefont
  {Arpaia}}, \bibinfo {author} {\bibfnamefont {S.}~\bibnamefont {Nawaz}},
  \bibinfo {author} {\bibfnamefont {F.}~\bibnamefont {Lombardi}}, \ and\
  \bibinfo {author} {\bibfnamefont {T.}~\bibnamefont {Bauch}},\ }\href
  {\doibase 10.1109/TASC.2013.2247454} {\bibfield  {journal} {\bibinfo
  {journal} {IEEE Trans. Appl. Supercond.}\ }\textbf {\bibinfo {volume} {23}},\
  \bibinfo {pages} {1101505} (\bibinfo {year} {2013})}\BibitemShut {NoStop}%
\bibitem [{\citenamefont {Baghdadi}\ \emph
  {et~al.}(2015{\natexlab{b}})\citenamefont {Baghdadi}, \citenamefont {Arpaia},
  \citenamefont {Charpentier}, \citenamefont {Golubev}, \citenamefont {Bauch},\
  and\ \citenamefont {Lombardi}}]{baghdadi2015fabricating}%
  \BibitemOpen
  \bibfield  {author} {\bibinfo {author} {\bibfnamefont {R.}~\bibnamefont
  {Baghdadi}}, \bibinfo {author} {\bibfnamefont {R.}~\bibnamefont {Arpaia}},
  \bibinfo {author} {\bibfnamefont {S.}~\bibnamefont {Charpentier}}, \bibinfo
  {author} {\bibfnamefont {D.}~\bibnamefont {Golubev}}, \bibinfo {author}
  {\bibfnamefont {T.}~\bibnamefont {Bauch}}, \ and\ \bibinfo {author}
  {\bibfnamefont {F.}~\bibnamefont {Lombardi}},\ }\href {\doibase
  10.1103/PhysRevApplied.4.014022} {\bibfield  {journal} {\bibinfo  {journal}
  {Phys. Rev. Applied}\ }\textbf {\bibinfo {volume} {4}},\ \bibinfo {pages}
  {014022} (\bibinfo {year} {2015}{\natexlab{b}})}\BibitemShut {NoStop}%
\bibitem [{\citenamefont {Nawaz}\ \emph {et~al.}(2013)\citenamefont {Nawaz},
  \citenamefont {Arpaia}, \citenamefont {Lombardi},\ and\ \citenamefont
  {Bauch}}]{nawaz2013microwave}%
  \BibitemOpen
  \bibfield  {author} {\bibinfo {author} {\bibfnamefont {S.}~\bibnamefont
  {Nawaz}}, \bibinfo {author} {\bibfnamefont {R.}~\bibnamefont {Arpaia}},
  \bibinfo {author} {\bibfnamefont {F.}~\bibnamefont {Lombardi}}, \ and\
  \bibinfo {author} {\bibfnamefont {T.}~\bibnamefont {Bauch}},\ }\href
  {\doibase 10.1103/PhysRevLett.110.167004} {\bibfield  {journal} {\bibinfo
  {journal} {Phys. Rev. Lett.}\ }\textbf {\bibinfo {volume} {110}},\ \bibinfo
  {pages} {167004} (\bibinfo {year} {2013})}\BibitemShut {NoStop}%
\bibitem [{\citenamefont {Arpaia}\ \emph
  {et~al.}(2014{\natexlab{a}})\citenamefont {Arpaia}, \citenamefont {Arzeo},
  \citenamefont {Nawaz}, \citenamefont {Charpentier}, \citenamefont
  {Lombardi},\ and\ \citenamefont {Bauch}}]{arpaia2014ultra}%
  \BibitemOpen
  \bibfield  {author} {\bibinfo {author} {\bibfnamefont {R.}~\bibnamefont
  {Arpaia}}, \bibinfo {author} {\bibfnamefont {M.}~\bibnamefont {Arzeo}},
  \bibinfo {author} {\bibfnamefont {S.}~\bibnamefont {Nawaz}}, \bibinfo
  {author} {\bibfnamefont {S.}~\bibnamefont {Charpentier}}, \bibinfo {author}
  {\bibfnamefont {F.}~\bibnamefont {Lombardi}}, \ and\ \bibinfo {author}
  {\bibfnamefont {T.}~\bibnamefont {Bauch}},\ }\href {\doibase
  10.1063/1.4866277} {\bibfield  {journal} {\bibinfo  {journal} {Appl. Phys.
  Lett.}\ }\textbf {\bibinfo {volume} {104}},\ \bibinfo {pages} {072603}
  (\bibinfo {year} {2014}{\natexlab{a}})}\BibitemShut {NoStop}%
\bibitem [{\citenamefont {Arzeo}\ \emph {et~al.}(2016)\citenamefont {Arzeo},
  \citenamefont {Arpaia}, \citenamefont {Baghdadi}, \citenamefont {Lombardi},\
  and\ \citenamefont {Bauch}}]{arzeo2016toward}%
  \BibitemOpen
  \bibfield  {author} {\bibinfo {author} {\bibfnamefont {M.}~\bibnamefont
  {Arzeo}}, \bibinfo {author} {\bibfnamefont {R.}~\bibnamefont {Arpaia}},
  \bibinfo {author} {\bibfnamefont {R.}~\bibnamefont {Baghdadi}}, \bibinfo
  {author} {\bibfnamefont {F.}~\bibnamefont {Lombardi}}, \ and\ \bibinfo
  {author} {\bibfnamefont {T.}~\bibnamefont {Bauch}},\ }\href {\doibase
  10.1063/1.4948477} {\bibfield  {journal} {\bibinfo  {journal} {J. Appl.
  Phys.}\ }\textbf {\bibinfo {volume} {119}},\ \bibinfo {pages} {174501}
  (\bibinfo {year} {2016})}\BibitemShut {NoStop}%
\bibitem [{\citenamefont {Arpaia}\ \emph {et~al.}(2016)\citenamefont {Arpaia},
  \citenamefont {Arzeo}, \citenamefont {Baghdadi}, \citenamefont {Trabaldo},
  \citenamefont {Lombardi},\ and\ \citenamefont {Bauch}}]{arpaia2016improved}%
  \BibitemOpen
  \bibfield  {author} {\bibinfo {author} {\bibfnamefont {R.}~\bibnamefont
  {Arpaia}}, \bibinfo {author} {\bibfnamefont {M.}~\bibnamefont {Arzeo}},
  \bibinfo {author} {\bibfnamefont {R.}~\bibnamefont {Baghdadi}}, \bibinfo
  {author} {\bibfnamefont {E.}~\bibnamefont {Trabaldo}}, \bibinfo {author}
  {\bibfnamefont {F.}~\bibnamefont {Lombardi}}, \ and\ \bibinfo {author}
  {\bibfnamefont {T.}~\bibnamefont {Bauch}},\ }\href {\doibase
  10.1088/0953-2048/30/1/014008} {\bibfield  {journal} {\bibinfo  {journal}
  {Supercond. Sci. Technol.}\ }\textbf {\bibinfo {volume} {30}},\ \bibinfo
  {pages} {014008} (\bibinfo {year} {2016})}\BibitemShut {NoStop}%
\bibitem [{\citenamefont {Trabaldo}\ \emph {et~al.}(2019)\citenamefont
  {Trabaldo}, \citenamefont {Arpaia}, \citenamefont {Arzeo}, \citenamefont
  {Andersson}, \citenamefont {Golubev}, \citenamefont {Lombardi},\ and\
  \citenamefont {Bauch}}]{trabaldo2019transport}%
  \BibitemOpen
  \bibfield  {author} {\bibinfo {author} {\bibfnamefont {E.}~\bibnamefont
  {Trabaldo}}, \bibinfo {author} {\bibfnamefont {R.}~\bibnamefont {Arpaia}},
  \bibinfo {author} {\bibfnamefont {M.}~\bibnamefont {Arzeo}}, \bibinfo
  {author} {\bibfnamefont {E.}~\bibnamefont {Andersson}}, \bibinfo {author}
  {\bibfnamefont {D.}~\bibnamefont {Golubev}}, \bibinfo {author} {\bibfnamefont
  {F.}~\bibnamefont {Lombardi}}, \ and\ \bibinfo {author} {\bibfnamefont
  {T.}~\bibnamefont {Bauch}},\ }\href@noop {} {\bibfield  {journal} {\bibinfo
  {journal} {Supercond. Sci. Technol.}\ }\textbf {\bibinfo {volume} {32}},\
  \bibinfo {pages} {073001} (\bibinfo {year} {2019})}\BibitemShut {NoStop}%
\bibitem [{\citenamefont {Arpaia}\ \emph
  {et~al.}(2014{\natexlab{b}})\citenamefont {Arpaia}, \citenamefont
  {Charpentier}, \citenamefont {Toskovic}, \citenamefont {Bauch},\ and\
  \citenamefont {Lombardi}}]{arpaia2014yba2cu3o7}%
  \BibitemOpen
  \bibfield  {author} {\bibinfo {author} {\bibfnamefont {R.}~\bibnamefont
  {Arpaia}}, \bibinfo {author} {\bibfnamefont {S.}~\bibnamefont {Charpentier}},
  \bibinfo {author} {\bibfnamefont {R.}~\bibnamefont {Toskovic}}, \bibinfo
  {author} {\bibfnamefont {T.}~\bibnamefont {Bauch}}, \ and\ \bibinfo {author}
  {\bibfnamefont {F.}~\bibnamefont {Lombardi}},\ }\href {\doibase
  10.1016/j.physc.2014.06.015} {\bibfield  {journal} {\bibinfo  {journal}
  {Physica C}\ }\textbf {\bibinfo {volume} {506}},\ \bibinfo {pages} {184}
  (\bibinfo {year} {2014}{\natexlab{b}})}\BibitemShut {NoStop}%
\bibitem [{\citenamefont {Ando}\ \emph {et~al.}(2002)\citenamefont {Ando},
  \citenamefont {Segawa}, \citenamefont {Komiya},\ and\ \citenamefont
  {Lavrov}}]{ando2002electrical}%
  \BibitemOpen
  \bibfield  {author} {\bibinfo {author} {\bibfnamefont {Y.}~\bibnamefont
  {Ando}}, \bibinfo {author} {\bibfnamefont {K.}~\bibnamefont {Segawa}},
  \bibinfo {author} {\bibfnamefont {S.}~\bibnamefont {Komiya}}, \ and\ \bibinfo
  {author} {\bibfnamefont {A.}~\bibnamefont {Lavrov}},\ }\href {\doibase
  10.1103/PhysRevLett.88.137005} {\bibfield  {journal} {\bibinfo  {journal}
  {Phys. Rev. Lett.}\ }\textbf {\bibinfo {volume} {88}},\ \bibinfo {pages}
  {137005} (\bibinfo {year} {2002})}\BibitemShut {NoStop}%
\bibitem [{Note3()}]{Note3}%
  \BibitemOpen
  \bibinfo {note} {The hole doping $p$, number of holes for planar copper atom,
  has been determined in our films indirectly, combining the knowledge of the
  $c$-axis parameter, estimated via XRD with that of the critical temperature
  $T_c$, determined via resistance vs temperature measurements. A detailed
  description of the analysis we have applied is presented in Ref. \cite
  {arpaia2018probing}.}\BibitemShut {Stop}%
\bibitem [{\citenamefont {Arpaia}\ \emph
  {et~al.}(2014{\natexlab{c}})\citenamefont {Arpaia}, \citenamefont {Golubev},
  \citenamefont {Baghdadi}, \citenamefont {Arzeo}, \citenamefont {Kunakova},
  \citenamefont {Charpentier}, \citenamefont {Nawaz}, \citenamefont
  {Lombardi},\ and\ \citenamefont {Bauch}}]{arpaia2014resistive}%
  \BibitemOpen
  \bibfield  {author} {\bibinfo {author} {\bibfnamefont {R.}~\bibnamefont
  {Arpaia}}, \bibinfo {author} {\bibfnamefont {D.}~\bibnamefont {Golubev}},
  \bibinfo {author} {\bibfnamefont {R.}~\bibnamefont {Baghdadi}}, \bibinfo
  {author} {\bibfnamefont {M.}~\bibnamefont {Arzeo}}, \bibinfo {author}
  {\bibfnamefont {G.}~\bibnamefont {Kunakova}}, \bibinfo {author}
  {\bibfnamefont {S.}~\bibnamefont {Charpentier}}, \bibinfo {author}
  {\bibfnamefont {S.}~\bibnamefont {Nawaz}}, \bibinfo {author} {\bibfnamefont
  {F.}~\bibnamefont {Lombardi}}, \ and\ \bibinfo {author} {\bibfnamefont
  {T.}~\bibnamefont {Bauch}},\ }\href {\doibase 10.1016/j.physc.2014.06.002}
  {\bibfield  {journal} {\bibinfo  {journal} {Physica C}\ }\textbf {\bibinfo
  {volume} {506}},\ \bibinfo {pages} {165} (\bibinfo {year}
  {2014}{\natexlab{c}})}\BibitemShut {NoStop}%
\bibitem [{\citenamefont {Charpentier}\ \emph {et~al.}(2016)\citenamefont
  {Charpentier}, \citenamefont {Arpaia}, \citenamefont {Gaudet}, \citenamefont
  {Matte}, \citenamefont {Baghdadi}, \citenamefont {L{\"o}fwander},
  \citenamefont {Golubev}, \citenamefont {Fournier}, \citenamefont {Bauch},\
  and\ \citenamefont {Lombardi}}]{charpentier2016hot}%
  \BibitemOpen
  \bibfield  {author} {\bibinfo {author} {\bibfnamefont {S.}~\bibnamefont
  {Charpentier}}, \bibinfo {author} {\bibfnamefont {R.}~\bibnamefont {Arpaia}},
  \bibinfo {author} {\bibfnamefont {J.}~\bibnamefont {Gaudet}}, \bibinfo
  {author} {\bibfnamefont {D.}~\bibnamefont {Matte}}, \bibinfo {author}
  {\bibfnamefont {R.}~\bibnamefont {Baghdadi}}, \bibinfo {author}
  {\bibfnamefont {T.}~\bibnamefont {L{\"o}fwander}}, \bibinfo {author}
  {\bibfnamefont {D.}~\bibnamefont {Golubev}}, \bibinfo {author} {\bibfnamefont
  {P.}~\bibnamefont {Fournier}}, \bibinfo {author} {\bibfnamefont
  {T.}~\bibnamefont {Bauch}}, \ and\ \bibinfo {author} {\bibfnamefont
  {F.}~\bibnamefont {Lombardi}},\ }\href@noop {} {\bibfield  {journal}
  {\bibinfo  {journal} {Phys. Rev. B}\ }\textbf {\bibinfo {volume} {94}},\
  \bibinfo {pages} {060503(R)} (\bibinfo {year} {2016})}\BibitemShut {NoStop}%
\bibitem [{\citenamefont {Baghdadi}\ \emph {et~al.}(2017)\citenamefont
  {Baghdadi}, \citenamefont {Arpaia}, \citenamefont {Stepantsov}, \citenamefont
  {Arzeo}, \citenamefont {Golubev}, \citenamefont {Montemurro}, \citenamefont
  {Andersson}, \citenamefont {Bauch},\ and\ \citenamefont
  {Lombardi}}]{baghdadi2017study}%
  \BibitemOpen
  \bibfield  {author} {\bibinfo {author} {\bibfnamefont {R.}~\bibnamefont
  {Baghdadi}}, \bibinfo {author} {\bibfnamefont {R.}~\bibnamefont {Arpaia}},
  \bibinfo {author} {\bibfnamefont {E.}~\bibnamefont {Stepantsov}}, \bibinfo
  {author} {\bibfnamefont {M.}~\bibnamefont {Arzeo}}, \bibinfo {author}
  {\bibfnamefont {D.}~\bibnamefont {Golubev}}, \bibinfo {author} {\bibfnamefont
  {D.}~\bibnamefont {Montemurro}}, \bibinfo {author} {\bibfnamefont
  {E.}~\bibnamefont {Andersson}}, \bibinfo {author} {\bibfnamefont
  {T.}~\bibnamefont {Bauch}}, \ and\ \bibinfo {author} {\bibfnamefont
  {F.}~\bibnamefont {Lombardi}},\ }\href@noop {} {\bibfield  {journal}
  {\bibinfo  {journal} {Phys. Rev. B}\ }\textbf {\bibinfo {volume} {95}},\
  \bibinfo {pages} {184505} (\bibinfo {year} {2017})}\BibitemShut {NoStop}%
\bibitem [{\citenamefont {Basov}\ \emph {et~al.}(1995)\citenamefont {Basov},
  \citenamefont {Liang}, \citenamefont {Bonn}, \citenamefont {Hardy},
  \citenamefont {Dabrowski}, \citenamefont {Quijada}, \citenamefont {Tanner},
  \citenamefont {Rice}, \citenamefont {Ginsberg},\ and\ \citenamefont
  {Timusk}}]{basov1995plane}%
  \BibitemOpen
  \bibfield  {author} {\bibinfo {author} {\bibfnamefont {D.}~\bibnamefont
  {Basov}}, \bibinfo {author} {\bibfnamefont {R.}~\bibnamefont {Liang}},
  \bibinfo {author} {\bibfnamefont {D.}~\bibnamefont {Bonn}}, \bibinfo {author}
  {\bibfnamefont {W.}~\bibnamefont {Hardy}}, \bibinfo {author} {\bibfnamefont
  {B.}~\bibnamefont {Dabrowski}}, \bibinfo {author} {\bibfnamefont
  {M.}~\bibnamefont {Quijada}}, \bibinfo {author} {\bibfnamefont
  {D.}~\bibnamefont {Tanner}}, \bibinfo {author} {\bibfnamefont
  {J.}~\bibnamefont {Rice}}, \bibinfo {author} {\bibfnamefont {D.}~\bibnamefont
  {Ginsberg}}, \ and\ \bibinfo {author} {\bibfnamefont {T.}~\bibnamefont
  {Timusk}},\ }\href {\doibase 10.1103/PhysRevLett.74.598} {\bibfield
  {journal} {\bibinfo  {journal} {Phys. Rev. Lett.}\ }\textbf {\bibinfo
  {volume} {74}},\ \bibinfo {pages} {598} (\bibinfo {year} {1995})}\BibitemShut
  {NoStop}%
\bibitem [{\citenamefont {Ishida}\ \emph {et~al.}(1997)\citenamefont {Ishida},
  \citenamefont {Okuda}, \citenamefont {Asaoka}, \citenamefont {Kazumata},
  \citenamefont {Noda},\ and\ \citenamefont {Takei}}]{ishida1997two}%
  \BibitemOpen
  \bibfield  {author} {\bibinfo {author} {\bibfnamefont {T.}~\bibnamefont
  {Ishida}}, \bibinfo {author} {\bibfnamefont {K.}~\bibnamefont {Okuda}},
  \bibinfo {author} {\bibfnamefont {H.}~\bibnamefont {Asaoka}}, \bibinfo
  {author} {\bibfnamefont {Y.}~\bibnamefont {Kazumata}}, \bibinfo {author}
  {\bibfnamefont {K.}~\bibnamefont {Noda}}, \ and\ \bibinfo {author}
  {\bibfnamefont {H.}~\bibnamefont {Takei}},\ }\href@noop {} {\bibfield
  {journal} {\bibinfo  {journal} {Phys. Rev. B}\ }\textbf {\bibinfo {volume}
  {56}},\ \bibinfo {pages} {11897} (\bibinfo {year} {1997})}\BibitemShut
  {NoStop}%
\bibitem [{\citenamefont {Arpaia}\ \emph {et~al.}(2019)\citenamefont {Arpaia},
  \citenamefont {Caprara}, \citenamefont {Fumagalli}, \citenamefont
  {De~Vecchi}, \citenamefont {Peng}, \citenamefont {Andersson}, \citenamefont
  {Betto}, \citenamefont {De~Luca}, \citenamefont {Brookes}, \citenamefont
  {Lombardi}, \citenamefont {Salluzzo}, \citenamefont {Braicovich},
  \citenamefont {Di~Castro}, \citenamefont {Grilli},\ and\ \citenamefont
  {Ghiringhelli}}]{arpaia2018dynamical}%
  \BibitemOpen
  \bibfield  {author} {\bibinfo {author} {\bibfnamefont {R.}~\bibnamefont
  {Arpaia}}, \bibinfo {author} {\bibfnamefont {S.}~\bibnamefont {Caprara}},
  \bibinfo {author} {\bibfnamefont {R.}~\bibnamefont {Fumagalli}}, \bibinfo
  {author} {\bibfnamefont {G.}~\bibnamefont {De~Vecchi}}, \bibinfo {author}
  {\bibfnamefont {Y.}~\bibnamefont {Peng}}, \bibinfo {author} {\bibfnamefont
  {E.}~\bibnamefont {Andersson}}, \bibinfo {author} {\bibfnamefont
  {D.}~\bibnamefont {Betto}}, \bibinfo {author} {\bibfnamefont
  {G.}~\bibnamefont {De~Luca}}, \bibinfo {author} {\bibfnamefont
  {N.}~\bibnamefont {Brookes}}, \bibinfo {author} {\bibfnamefont
  {F.}~\bibnamefont {Lombardi}}, \bibinfo {author} {\bibfnamefont
  {M.}~\bibnamefont {Salluzzo}}, \bibinfo {author} {\bibfnamefont
  {L.}~\bibnamefont {Braicovich}}, \bibinfo {author} {\bibfnamefont
  {C.}~\bibnamefont {Di~Castro}}, \bibinfo {author} {\bibfnamefont
  {M.}~\bibnamefont {Grilli}}, \ and\ \bibinfo {author} {\bibfnamefont
  {G.}~\bibnamefont {Ghiringhelli}},\ }\href {\doibase 10.1126/science.aav1315}
  {\bibfield  {journal} {\bibinfo  {journal} {Science}\ }\textbf {\bibinfo
  {volume} {365}},\ \bibinfo {pages} {906} (\bibinfo {year}
  {2019})}\BibitemShut {NoStop}%
  \bibitem [{\citenamefont {Scherer}\ \emph {et~al.}(1992)\citenamefont
  {Scherer}, \citenamefont {Marienhoff}, \citenamefont {Herwig}, \citenamefont
  {Neuhaus},\ and\ \citenamefont {Jutzi}}]{scherer1992anisotropy}%
  \BibitemOpen
  \bibfield  {author} {\bibinfo {author} {\bibfnamefont {T.}~\bibnamefont
  {Scherer}}, \bibinfo {author} {\bibfnamefont {P.}~\bibnamefont {Marienhoff}},
  \bibinfo {author} {\bibfnamefont {R.}~\bibnamefont {Herwig}}, \bibinfo
  {author} {\bibfnamefont {M.}~\bibnamefont {Neuhaus}}, \ and\ \bibinfo
  {author} {\bibfnamefont {W.}~\bibnamefont {Jutzi}},\ }\href {\doibase
  10.1016/0921-4534(92)90238-8} {\bibfield  {journal} {\bibinfo  {journal}
  {Physica C}\ }\textbf {\bibinfo {volume} {197}},\ \bibinfo {pages} {79}
  (\bibinfo {year} {1992})}\BibitemShut {NoStop}%
\bibitem [{\citenamefont {Klemenz}\ \emph {et~al.}(1999)\citenamefont
  {Klemenz}, \citenamefont {Utke},\ and\ \citenamefont
  {Scheel}}]{klemenz1999defects}%
  \BibitemOpen
  \bibfield  {author} {\bibinfo {author} {\bibfnamefont {C.}~\bibnamefont
  {Klemenz}}, \bibinfo {author} {\bibfnamefont {I.}~\bibnamefont {Utke}}, \
  and\ \bibinfo {author} {\bibfnamefont {H.}~\bibnamefont {Scheel}},\ }\href
  {\doibase 10.1016/S0022-0248(99)00356-5} {\bibfield  {journal} {\bibinfo
  {journal} {J. Cryst. Growth}\ }\textbf {\bibinfo {volume} {207}},\ \bibinfo
  {pages} {62} (\bibinfo {year} {1999})}\BibitemShut {NoStop}%
\bibitem [{\citenamefont {Khanof}\ \emph {et~al.}(2018)\citenamefont {Khanof},
  \citenamefont {Mannhart},\ and\ \citenamefont
  {Boschker}}]{khanof2018coherently}%
  \BibitemOpen
  \bibfield  {author} {\bibinfo {author} {\bibfnamefont {S.}~\bibnamefont
  {Khanof}}, \bibinfo {author} {\bibfnamefont {J.}~\bibnamefont {Mannhart}}, \
  and\ \bibinfo {author} {\bibfnamefont {H.}~\bibnamefont {Boschker}},\ }\href
  {\doibase 10.1063/1.5037540} {\bibfield  {journal} {\bibinfo  {journal}
  {Appl. Phys. Lett.}\ }\textbf {\bibinfo {volume} {113}},\ \bibinfo {pages}
  {022605} (\bibinfo {year} {2018})}\BibitemShut {NoStop}%
\bibitem [{\citenamefont {Ye}\ and\ \citenamefont
  {Nakamura}(1993)}]{ye1993quantitative}%
  \BibitemOpen
  \bibfield  {author} {\bibinfo {author} {\bibfnamefont {J.}~\bibnamefont
  {Ye}}\ and\ \bibinfo {author} {\bibfnamefont {K.}~\bibnamefont {Nakamura}},\
  }\href {\doibase 10.1103/PhysRevB.48.7554} {\bibfield  {journal} {\bibinfo
  {journal} {Phys. Rev. B}\ }\textbf {\bibinfo {volume} {48}},\ \bibinfo
  {pages} {7554} (\bibinfo {year} {1993})}\BibitemShut {NoStop}%
\end{thebibliography}%

\end{document}